\documentclass[a4paper,fleqn,usenatbib]{mnras}

\usepackage{newtxtext,newtxmath}
\usepackage[T1]{fontenc}
\usepackage{ae,aecompl}

\usepackage{graphicx}	
\usepackage{multicol}
\usepackage{amsmath}	
\usepackage[nolist]{acronym}
\usepackage{upgreek}
\usepackage[separate-uncertainty=true,multi-part-units=brackets]{siunitx}
\usepackage[flushleft]{threeparttable}

\title[Structures of $z\approx 3$ quiescent galaxies]{Compact, bulge dominated structures of spectroscopically confirmed quiescent galaxies at $z\approx 3$}
\author[P. Lustig et al.]{Peter Lustig,$^{1}$\thanks{E-mail: peter.lustig@physik.lmu.de}
Veronica Strazzullo,$^{1,2,3,4}$
Chiara D'Eugenio,$^{5}$
Emanuele Daddi,$^{5}$\newauthor
Maurilio Pannella,$^{1,2}$
Alvio Renzini,$^{6}$
Andrea Cimatti,$^{7,8}$
Raphael Gobat,$^{9}$\newauthor
Shuowen Jin,$^{10,11}$
Joseph J. Mohr$^{1,12}$
and Masato Onodera$^{13,14}$
\\
$^{1}$Faculty of Physics, Ludwig-Maximilians-Universität, Scheinerstr. 1, 81679 Munich, Germany\\
$^{2}$University of Trieste, Piazzale Europa, 1, 34127 Trieste TS, Italy\\
$^{3}$INAF–Osservatorio Astronomico di Brera, Via Brera 28, 20121 Milano, Italy\\
$^{4}$INAF - Osservatorio Astronomico di Trieste, via Tiepolo 11, I-34131, Trieste, Italy\\
$^{5}$CEA, IRFU, DAp, AIM, Université Paris-Saclay, Université Paris Diderot, Sorbonne Paris Cit, CNRS, F-91191 Gif-sur-Yvette, France\\
$^{6}$INAF - Osservatorio Astronomico di Padova, Vicolo dell'Osservatorio 5, I-35122 Padova, Italy\\
$^{7}$University of Bologna, Department of Physics and Astronomy (DIFA), Via Gobetti 93/2, I-40129, Bologna, Italy\\
$^{8}$INAF - Osservatorio Astrofisico di Arcetri, Largo E. Fermi 5, I-50125, Firenze, Italy \\
$^{9}$Instituto de Física, Pontificia Universidad Católica de Valparaíso, Casilla, 4059, Valparaíso, Chile \\
$^{10}$Instituto de Astrofísica de Canarias (IAC), E-38205 La Laguna, Tenerife, Spain\\
$^{11}$Universidad de La Laguna, Dpto. Astrofísica, E-38206 La Laguna,Tenerife, Spain\\
$^{12}$Max Planck Institute for Extraterrestrial Physics, Giessenbachstrasse, 85748 Garching, Germany\\
$^{13}$Subaru Telescope, National Astronomical Observatory of Japan, National Institutes of Natural Sciences, 650 North A’ohoku Place, Hilo, HI 96720, USA \\
$^{14}$Department of Astronomical Science, The Graduate University for Advanced Studies, SOKENDAI, 2-21-1 Osawa, Mitaka, Tokyo, 181-8588, Japan
}

\date{Accepted XXX. Received YYY; in original form ZZZ}

\pubyear{2020}

\begin{document}

\newcommand{\sersic}{Sérsic}
\newcommand{\re}{r_{\mathrm{e}}}
\newcommand{\Mstar}{M_{\star}}
\newcommand{\Msol}{M_{\odot}}
\newcommand{\lmass}{\log(\Mstar/\Msol)}
\newcommand{\Hmag}{H_{\textrm{AB}}}
\newcommand{\asunc}[3]{#1_{-#2}^{+#3}}
\newcommand{\asuncunit}[4]{#1_{-#2}^{+#3}\,\textrm{#4}}
\newcommand{\lobs}{\lambda_{\mathrm{obs}}}
\newcommand{\dlrdll}{\frac{\Delta\log \re}{\Delta\log \lambda}}
\newcommand{\bzk}{BzK }
\newcommand{\zphot}{z_{\textrm{phot}}}
\newcommand{\zspec}{z_{\textrm{spec}}}
\newcommand{\lrho}{\log(\rho_1\,\textrm{kpc}^3/\Msol)}
\newcommand{\vc}{v_{1}}

\begin{acronym}
	\acro{agn}[AGN]{active galaxy nucleus}
	\acroplural{agn}[AGNs]{active galaxy nuclei}
	\acro{et}[ET]{early-type}
	\acro{lt}[LT]{late-type}
   \acro{fwhm}[FWHM]{full width at half maximum}
	\acro{hst}[HST]{Hubble Space Telescope}
	\acro{imf}[IMF]{initial mass function}
	\acro{mtl}[$M/L$]{mass-to-light}
	\acro{nir}[NIR]{near-infrared}
	\acro{pbzk}[pBzK]{passive BzK}
	\acro{photoz}[photo-z]{photometric redshift}
	\acro{psb}[PSB]{post-starburst}
	\acro{psf}[PSF]{point spread function}
	\acro{rms}[RMS]{root mean square}
	\acro{sed}[SED]{spectral energy distribution}
	\acro{sfh}[SFH]{star formation history}
	\acroplural{sfh}[SFHs]{star formation histories}
	\acro{sfr}[SFR]{star formation rate}
	\acro{snr}[SNR]{signal-to-noise ratio}
	\acro{ssfr}[sSFR]{specific star formation rate}
	\acro{wfc3}[WFC3]{Wide Field Camera 3}
\end{acronym}

\label{firstpage}

\pagerange{\pageref{firstpage}--\pageref{lastpage}}
\maketitle

\begin{abstract}
We study structural properties of spectroscopically confirmed massive quiescent galaxies at $z\approx3$ with one of the first sizeable samples of such sources, made of ten $10.8<\lmass<11.3$ galaxies at $2.4 < z < 3.2$ in the COSMOS field whose redshifts and quiescence are confirmed by HST grism spectroscopy.
Although affected by a weak bias toward younger stellar populations, this sample is deemed to be largely representative of the majority of the most massive and thus intrinsically rarest quiescent sources at this cosmic time. We rely on targeted HST/WFC3 observations and fit \sersic\ profiles to the galaxy surface brightness distributions at $\approx\SI{4000}{\angstrom}$ restframe. We find typically high \sersic\ indices and axis ratios (medians $\approx 4.5$ and $0.73$, respectively) suggesting that, at odds with some previous results, the first massive quiescent galaxies may largely be already bulge-dominated systems. We measure compact galaxy sizes with an average of $\approx\SI{1.4}{kpc}$ at $\lmass\approx 11.2$, in good agreement with the extrapolation at the highest masses of previous determinations of the stellar mass - size relation of quiescent galaxies, and of its redshift evolution, from photometrically selected samples at lower and similar redshifts. This work confirms the existence of a population of compact, bulge dominated, massive, quiescent sources at $z\approx 3$, providing one of the first statistical estimates of their structural properties, and further constraining the early formation and evolution of the first quiescent galaxies.

\end{abstract}

\begin{keywords}
 galaxies: evolution -- galaxies: high-redshift -- galaxies: structure
\end{keywords}


\section{Introduction}

Structural properties of galaxies in the nearby universe correlate with their stellar population properties. Early-type galaxies are characterised by a higher central concentration and typically lower apparent ellipticity than late-type galaxies, and generally have a low \ac{ssfr}. 
Up to a stellar mass of $\lmass\approx 11$ early-type galaxies are more compact than late-type galaxies and show a steeper stellar mass vs. size relation \citep[e.g.,][]{Shen_2003, Guo_2009}.

Up to $z\approx 1$, high axis ratios are largely ubiquitous in the most massive $\lmass\gtrsim 11$ quiescent galaxies, although larger fractions of lower mass galaxies show lower axis ratios; this suggests that the mechanisms forming the most massive quiescent sources also result in the formation of bulge-dominated, spheroidal structures \citep{VanDerWel_2009, Holden_2012}. In fact, integral field spectroscopy showed that the vast majority of early-type galaxies in the nearby universe are fast rotators, with slow rotators dominating the early-type galaxy population only at the high mass end \citep[$\Mstar\gtrsim 2\times10^{11}\Msol$; e.g.,][]{Emsellem_2011, Cappellari_2016}.

Structural properties of massive galaxies at higher redshift are more sparsely investigated and have produced more controversial results.
\cite{Stockton_2004, Stockton_2008} provided the first constraints on the structure of massive quiescent galaxies at $z\approx2.5$ and revealed a higher fraction of quiescent galaxies with low \sersic\ index profiles and smaller axis ratios with respect to low-redshift samples. Such scenario has been strengthened by following works with larger samples \citep{VanDokkum_2008, McGrath_2008, Bundy_2010, VanDerWel_2011, Chang_2013, McLure_2013, Hsu_2014, Bezanson_2018}.
Recently \cite{Hill_2019} investigated the axis ratio evolution of star-forming and quiescent galaxies over the redshift range $0.2<z<4.0$, finding that massive ($\lmass>11$) quiescent galaxies at $2.5<z<3.5$ are as flat as star-forming galaxies. Limited measurements of rotation curves indeed provide evidence for the existence of rotationally supported massive quiescent galaxies at high redshift \citep{Newman_2015, Newman_2018, Toft_2017}.
Nonetheless, the coupling of structural and stellar population properties of galaxies at higher redshifts remains debated, as other studies find that the correlation between early-type structure and low \ac{ssfr} holds at least up to $z\approx 3$, suggesting that morphological transformation towards bulge-dominated systems is tightly related to quenching of star formation already at high redshift \citep{Bell_2012, Lang_2014, Tacchella_2015, Mowla_2019_b, Esdaile_2020}.
It has in fact been shown that galaxies beyond a given stellar mass or central stellar mass density threshold are largely quiescent \citep[e.g.,][]{Kauffmann_2003, Brinchmann_2004, Franx_2008, Peng_2010, Dokkum_2015, Whitaker_2017}, and that - although it remains unclear whether mass or density is the actual driver \citep{Lilly_2016} - 
the most massive star-forming galaxies that at high redshift approach such density threshold are very likely to rapidly quench, given the drop in their number density at lower redshifts \citep{Mowla_2019_b}.

A possible mechanism to explain the correlation between structural and stellar population properties is the compaction of a star-forming disk in a first step, followed by quenching (possibly also as a consequence of the morphological transformation). The compaction of the disk can be a result of gas inflow from filaments or mergers \citep[e.g.,][]{Birnboim_2003, Keres_2005, Dekel_2006, Dekel_2009a}, causing violent disk instabilities that drive dissipative gas inflow in the center. This leads to a compact galaxy with a high star formation rate \citep[e.g.,][]{Dekel_2009b, Burkert_2010, Dekel_2013, Dekel_2014, Zolotov_2015, Gomez-Guijarro_2019, Wu_2020}.
Multiple mechanisms can then quench star formation, as suggested by simulations: gas consumption by star formation, stellar and \ac{agn} feedback as well as morphological quenching can produce fast quenching at high redshift, while virial shock heating, gravitational infall and \ac{agn} feedback can maintain quenching at lower redshift \citep[e.g.,][]{Dekel_1986, Birnboim_2003, Keres_2005, Dekel_2006, Ciotti_2007, Dekel_2008, Khochfar_2008, Martig_2009, Dekel_2009a, Tachella_2016}.
Bulges embedded in star-forming disks can remain starved from accreted gas and maintain quenching if the infalling gas has a too high angular momentum to reach the bulge \citep{Renzini_2018}.

Many studies have shown that the average size of distant quiescent galaxies at a given stellar mass is lower than for lower-redshift counterparts \citep[e.g.,][among many others]{Daddi_2005, Trujillo_2006, Toft_2007, Cimatti_2008, Cimatti_2012, Carollo_2013, Cassata_2013, Kubo2018, Mowla_2019_b}. Although an evolution in the average size at fixed mass is also observed for late-type galaxies, it is milder than for early-types.
With a large sample drawn from the CANDELS/3D-HST survey \citep{Grogin2011_CandelsDesign, Koekemoer2011_CandelsData, Momcheva2016_3Dhst}, \cite{VanDerWel2014} studied morphologies of quiescent and star-forming galaxies with redshifts $0<z<3$.
They find a redshift independent slope of the mass-size relation that is steeper for quiescent than for star-forming galaxies, and a size growth of massive quiescent galaxies of nearly an order of magnitude since $z\approx 3$, compared to a factor $\approx 3$ for star-forming sources.
Using measurements from the COSMOS-DASH survey,  \cite{Mowla_2019_b} extended the \cite{VanDerWel2014} sample to higher stellar masses (162 galaxies at $1.5<z<3.0$ with $\lmass>11.3$), which are poorly probed in the CANDELS/3D-HST survey due to the intrinsically very low number density of such sources, and find consistent results. However, even this survey only adds two quiescent galaxies to the \cite{VanDerWel2014} sample at $z>2.5$.
As an alternative to overcome the problem of small sample sizes of the most distant, massive quiescent galaxies in deep fields, targeted imaging has been used to study these objects up to $z\approx 4$ \citep{Straatman2015, Kubo2018}, supporting the findings of strong average size growth of the quiescent galaxy population.

The observed redshift evolution of the mass-size relation of quiescent galaxies can be explained by a combination of different effects.
Although gas rich mergers, resulting in central starbursts, are not an efficient way to increase galaxy size \citep{Lin_2007, Lin_2008, Perez_2011, Athanassoula_2016}, gas poor minor mergers are often considered a viable and potentially significant channel for size growth of quiescent galaxies \citep{Khochfar_2006, Bell_2006, Naab_2006, Naab_2009, Lin_2008,  Bezanson_2009, Oser_2010, Oser_2012, Trujillo_2011, Bedorf_2013}.
Progenitor bias is also often considered as an important contribution to the evolution of the mass-size relation, because of the significant drop of the quiescent galaxy population towards higher redshifts, implying a progenitor-descendant mismatch when comparing quiescent galaxy samples at different redshifts \citep{Dokkum_1996, Dokkum_2001, Poggianti_2013, Carollo_2013, Cassata_2013}.
To minimise the effect of progenitor bias, \cite{Belli_2014} and \cite{Stockmann_2020} investigated size evolution at constant velocity dispersion \citep[which is found to remain approximately unchanged for quiescent systems,][]{VanDerWel_2009_a, Bezanson_2012}, finding that size growth of individual galaxies may in fact have a significant role in the observed mass-size evolution. The observed evolution is thus likely produced by a combination of both galaxy growth and progenitor bias.
Additional complications come from the use of light as a tracer of stellar mass. Radial color -- and thus mass-to-light ratio -- gradients can lead to significant differences between half-light and half-mass radii. \cite{Suess_2019, Suess_2019b} find that color gradients of quiescent galaxies are nearly flat at $z\gtrsim 2$, increase with decreasing redshift and are stronger in massive, larger and redder galaxies.
Stellar mass vs. half-mass size relations of quiescent galaxies are shallower than stellar mass vs. (restframe optical) half-light size relations, and the growth of half-mass sizes towards lower redshifts is milder than for optical half-light sizes.

In most studies of the highest redshift quiescent sources, relying on purely photometric observations, the classification of star-forming vs. quiescent galaxies is performed by exploiting the correlation between \ac{ssfr} and galaxy colors in properly chosen passbands \citep{Daddi2004, Labbe_2005, Williams_2009, Ilbert_2010}.
 Especially at high redshift, where the number density of massive quiescent galaxies and the quiescent galaxy fraction decrease significantly \citep{Whitaker_2010, Marchesini_2010, Brammer_2011, Ilbert_2013, Muzzin_2013b, Mowla_2019_b} and the bimodality in color sequences
 is less pronounced \citep{Muzzin_2013, LaigleCOSMOS2016}, misclassification can lead to a significant contamination of quiescent galaxy samples from star-forming objects.
 Spectroscopic confirmation of quiescence can help securing higher-purity samples of quiescent galaxies. However, spectroscopically confirming very distant quiescent sources is difficult and observationally expensive compared to star-forming galaxies at similar redshifts because of the lack of strong emission lines.
Direct spectroscopic confirmation of quiescent sources currently reaches out to $z\approx 4$, and is based on the $\SI{4000}{\angstrom}$ break, overall continuum shape, and/or weaker features as Fe and Mg absorption lines \citep{Glazebrook_2004, Cimatti_2004, Kriek_2006, Gobat_2012, Onodera_2012, Onodera_2015, Newman_2015, Marsan_2015, Hill_2016, Glazebrook_2017, Marsan_2017, Gobat_2017, Newman_2018, Schreiber_2018, Tanaka_2019, Forrest_2020, Forrest_2020b, Valentino_2020, Esdaile_2020}.
The morphological properties of sizeable samples of spectroscopically confirmed quiescent galaxies have only been analysed up to $z<2.3$ \citep{Cimatti_2008, VanDokkum_2008, Belli_2017, Stockmann_2020}. At higher redshifts investigations are limited to a handful of galaxies at most \citep{Gobat_2012, Marsan_2015, Hill_2016, Tanaka_2019, Esdaile_2020}.
In this work we investigate structural properties of a spectroscopically confirmed sample of 10 quiescent galaxies at $2.4<z<3.2$ with stellar masses of $\lmass\gtrsim 11$, relying on targeted \ac{hst} WFC3/F160W imaging and G141 grism observations. This sample contains $\approx 1/4$ of all spectroscopically confirmed quiescent galaxies at $z>2.4$.
Our sample is presented in Section~\ref{sec:sample}. In Section~\ref{sec:morphology_fit} we explain our analysis and methods. In Section~\ref{sec:results} we present and discuss our results. Section~\ref{sec:summary} summarizes our findings and conclusions.

We assume a $\Lambda$CDM cosmology with $H_0=71$, $\Omega_{\textrm{M}}=0.27$ and $\Omega_{\Lambda}=0.73$. Magnitudes are given in the AB system.

\section{The quiescent galaxy sample}
\label{sec:sample}
\subsection{Sample selection}
\label{sec:sample_selection}

\begin{figure}
	\includegraphics[width=\columnwidth]{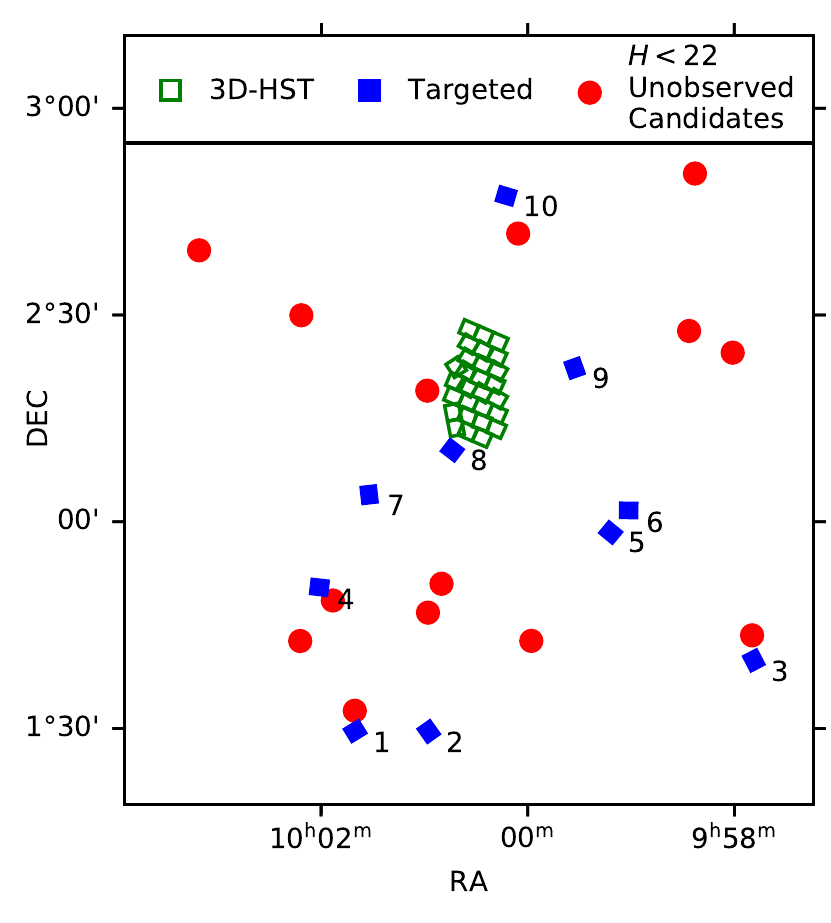}
	\vspace*{-4mm}
	\caption{Selection of targets for HST follow-up observations in the $\SI{2}{deg^2}$ COSMOS field. Blue rectangles show the WFC3 footprint of the observations acquired for this project. Red symbols show the remaining quiescent galaxies from our $H<22$ , $z>2.5$ candidate sample that were not included in the target list. Green rectangles show the footprint of the 3D-HST survey field in COSMOS.}
	\label{fig:observed_sky}
\end{figure}

We selected high-redshift quiescent galaxy candidates for \ac{hst} grism follow-up from the \cite{McCracken_2010} photometric catalog of the $\SI{2}{deg^2}$ COSMOS field. Initially we selected \ac{pbzk} galaxies \citep{Daddi2004} that satisfy
the conditions:
\begin{eqnarray}
\mathrm{BzK} =& (z_{\textrm{AB}}-K_{\textrm{AB}}) - (B_{\textrm{AB}}-z_{\textrm{AB}}) &>-0.2\\
\mathrm{zK} =&  (z_{\textrm{AB}}-K_{\textrm{AB}}) &> 2.5.
\label{eq:BzK}
\end{eqnarray}
These criteria select high-redshift (typically $z \gtrsim 1.4$) passive galaxies purely based on observed colors, without relying on photometric redshift estimation or \ac{sed} analysis to identify quiescent sources.
Given the low \ac{snr} of even massive quiescent galaxies at $z>2$ in the available $B$ and $z$ band imaging, leading to large uncertainties in the formal passive vs. star-forming BzK classification of such sources, we also retained galaxies with $\ac{snr}<5$ in these bands, independent of their classification as quiescent or star-forming BzK galaxies.
We then considered photometric redshifts ($\zphot$) for the selected galaxies estimated with the software EAZY \citep{Brammer_2008} and specifically calibrated to better estimate photometric redshifts of high-redshift quiescent galaxies \citep[see details in][]{Strazzullo_2015}. These photometric redshifts are listed in Table~\ref{tab:fitresults}. We removed from the sample all galaxies with $\zphot<2.5$, as well as galaxies classified as star-forming from their restframe UVJ colors \citep{Williams_2009} as estimated by EAZY assuming the galaxy photometric redshift.
We performed \ac{sed} fitting with FAST \citep{KriekFAST2009} with different model libraries, including 1) a generic setup with delayed exponential \ac{sfh} and dust attenuation up to $A_{\textrm{V}}=\SI{5}{mag}$, assuming a \cite{Calzetti2001} dust attenuation law, 2) constant \ac{sfr} with $A_{\textrm{V}}$ up to $5$ mag, 3) only quiescent (including very young quiescent, given the redshift of our targets) models ($\textrm{age}/\tau >4$, $\textrm{age}>\SI{0.5}{Gyr}$). Based on this analysis, we discarded all candidates with an \ac{sed} suggesting a possible star-forming solution.
To further reduce potential contamination of the \ac{hst} follow-up target sample from star-forming sources, we also deprioritised candidates with a \ac{snr} $\geq 4$ at $\SI{24}{\micro\metre}$ in the \cite{Le_Floc_h_2009} catalog, except if they had a well probed, convincingly quiescent SED with no plausible star-forming solution, suggesting that the $\SI{24}{\micro\metre}$ emission could be powered by an \ac{agn}.
We then selected from these candidates suitable targets for \ac{hst} grism follow-up observations. In order to observe a first, sizeable sample of $z\approx 3$ quiescent candidates, we focused on massive
galaxies for which a sufficiently high SNR spectrum to measure a reliable redshift could be obtained in $1-2$ orbits. To this aim, we simulated for each candidate the grism spectrum that could be obtained within this observing time, assuming the source photometric redshift and best-fit \ac{sed} model, modelling the simulated spectrum to estimate the redshift. This observational constraint largely limited the viable targets to sources brighter than $\Hmag\approx22$, leading to a sample of 23 sources that are shown in Figure~\ref{fig:observed_sky}.
Owing to the low number density of such massive, quiescent galaxies at $z\approx 3$, none of these objects is found in the CANDELS/3D-HST COSMOS field \citep{Grogin2011_CandelsDesign, Koekemoer2011_CandelsData, Momcheva2016_3Dhst}.
Although such bright ($\Hmag < 22$) targets were favoured because of the observational reasons discussed above, as well as of higher \ac{snr} photometry resulting in a more robust characterization of the galaxy \ac{sed}, we also explored fainter candidates that potentially allow us to probe higher redshift galaxies. We thus included in the final target sample a fainter ($\Hmag\approx 23$) source at $\zphot=3.2$ for which - in contrast to most similarly faint candidates - the \ac{sed} modeling discussed above was able to reject star-forming solutions at high confidence. The final target sample of 10 sources with photometric redshifts between 2.5 and 3.2 is listed in Table~\ref{tab:fitresults} and shown in Figure~\ref{fig:observed_sky}.

The selected targets have been observed with the G141 grism and direct imaging in the F160W band with the \ac{wfc3} on board of \ac{hst} (program ID 15229, PI: E. Daddi).
D'Eugenio et al. \citep[in preparation, see also][]{Deugenio_2020} have estimated spectroscopic redshifts from the grism spectra, which are shown in Table~\ref{tab:fitresults}. They combined the grism spectra with photometric measurements from the \cite{LaigleCOSMOS2016} catalog and performed a stellar population analysis. By comparing the goodness-of-fit of constant \ac{sfr} vs. passive templates with exponentially declining \acp{sfh},
star-forming solutions could be rejected for all galaxies in the sample (see full details and discussion in D'Eugenio et al. in preparation).

From the analysis of the stacked spectrum of all galaxies in the sample (except ID 7) \cite{Deugenio_2020} derived a \ac{ssfr} of $\SI{4.35\pm2.47 e-11}{yr^{-1}}$, which is 60 times below the main sequence of star-forming galaxies at the median redshift of $z=2.8$ \citep{Schreiber_2015}. The lookback time where $\SI{50}{percent}$ of the stellar mass of the stacked sample was formed is $t_{50}=\asuncunit{300}{50}{200}{Myr}$.

We note that the median and NMAD (normalised mean absolute difference) scatter of $(\zspec-\zphot)/(1+\zspec)$ for the sample studied here, using the grism redshifts from \cite{Deugenio_2020} and the \cite{Strazzullo_2015} photometric redshifts used for the sample selection, are 0.03 and 0.06; we thus assume that no significant biases are introduced in the sample studied here by uncertainties in the photometric redshifts used for the sample selection.

\subsection{SED modeling and stellar mass estimates}
\label{sec:sedfitting}
\begin{figure*}
	\includegraphics[width=2\columnwidth]{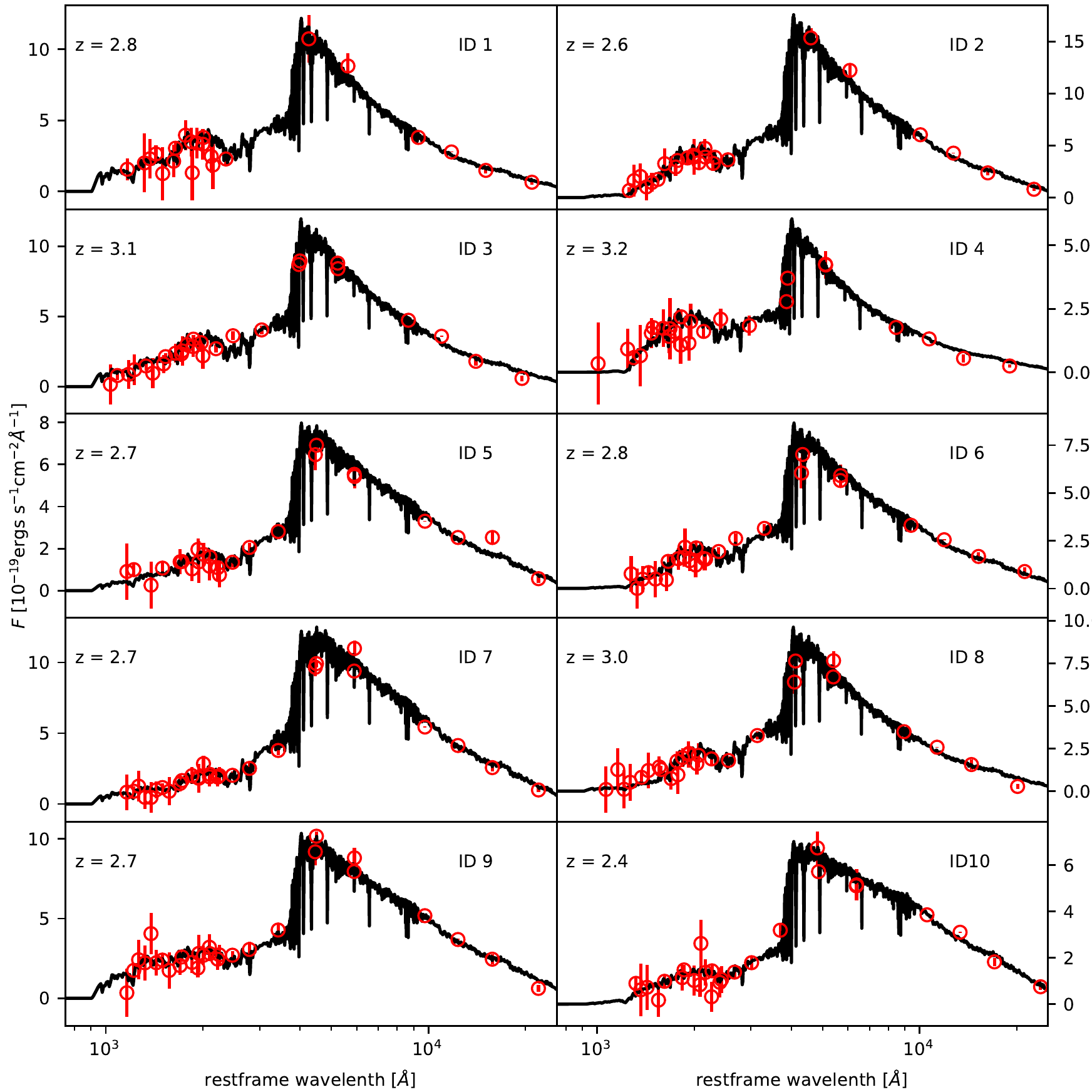}
	\caption{Observed \acp{sed} from $\SI{0.42}{\micro\metre}$ to $\SI{8}{\micro\metre}$ \protect\citep{LaigleCOSMOS2016} and best fit stellar population models (see Section~\ref{sec:sedfitting}).}
	\label{fig:sedfits}
\end{figure*}

We perform \ac{sed} fitting to estimate stellar masses of the targets from multi-band photometry from the COSMOS2015 catalog \citep[][]{LaigleCOSMOS2016} adopting the spectroscopic redshifts measured in \cite{Deugenio_2020}. We use FAST++\footnote{https://github.com/cschreib/fastpp} to fit \cite{Bruzual2003} population synthesis models to $29$ photometric bands\footnote{For sources that are observed in the $H$ and $Ks$ band by both UltraVISTA and WIRCam we have checked that there is no impact on the stellar mass estimates if the shallower WIRCAM data is removed.} from $\SI{0.42}{\micro\metre}$ to $\SI{8}{\micro\metre}$ (including narrow bands). We assume a \cite{Chabrier2003} \ac{imf}, a \cite{Calzetti2001} dust attenuation law and a delayed exponentially declining \ac{sfh} with $7\le\log(\tau/\textrm{yr})\le10$.
To allow for a more direct comparison with \citet[][see Section~\ref{sec:results}]{VanDerWel2014} we also estimate stellar masses assuming an exponentially declining \ac{sfh}, finding no systematics and individual stellar mass estimates differing by at most $\SI{0.05}{dex}$ for this specific sample, having no impact on our analysis. The metallicity is fixed to solar; leaving it free affects the mass estimates by at most $\SI{0.07}{dex}$. The best fit \ac{sed} models are shown in Figure \ref{fig:sedfits}. The formal uncertainties on the estimated stellar masses with the given \ac{sed} fitting setup are $\lesssim\SI{0.06}{dex}$; we stress that these uncertainties do not include known sources of statistical and systematic errors \citep[e.g.,][]{Maraston_2006, Longhetti_2009, Muzzin_2009, Conroy_2013, Pacifici_2015}, and that more realistic absolute uncertainties on the individual mass estimates are likely around a factor $\approx 2$.

IDs 2, 4, 7 and 10 have close neighbours in our F160W imaging that are undetected in the \cite{LaigleCOSMOS2016} catalog (see Section~\ref{sec:morphology_fit}). For these targets we scale the estimated stellar masses by the fraction of the target flux to the total flux including the undetected neighbours within the $\SI{3}{arcsec}$ aperture used in \cite{LaigleCOSMOS2016}, assuming the F160W fluxes measured in Section~\ref{sec:morphology_fit}. This correction decreases the masses of IDs 2, 4, 7 and 10 by $0.01, 0.11, 0.02$ and $\SI{0.01}{dex}$, respectively.  The resulting stellar masses are listed in Table~\ref{tab:fitresults}\footnote{We note that these masses reflect the total fluxes reported in the \cite{LaigleCOSMOS2016} catalog. We have verified that the total flux estimated by GALFIT on the F160W imaging is fully consistent with the total flux in the $H$ band from the \cite{LaigleCOSMOS2016} catalog (the average flux ratio for these targets is 0.96 with a dispersion of $\approx 0.15$, accounting for the small color term between the two filters).}.

The median estimated stellar mass of our sample is $\lmass = 11.16$ with individual masses in the range $10.8<\lmass<11.3$.
To ensure that no systematics affect our comparisons with \citet[][see Section~\ref{sec:results}]{VanDerWel2014}, who use stellar mass estimates from \cite{Skelton2014}, we estimate stellar masses with the same setup for sources from the \cite{Skelton2014} catalog using \cite{LaigleCOSMOS2016} photometry and redshifts from \cite{Skelton2014}. By comparison of the two estimates we find a statistical scatter on the estimated stellar masses of $\SI{0.1}{dex}$ and no systematics\footnote{We do not use stellar mass estimates from \protect\cite{LaigleCOSMOS2016} because: 1) we re-estimate stellar masses adopting the grism redshift (see also related discussion in Section~\ref{sec:representativeness} and Appendix~\ref{sec:specz_photoz}), and 2) as also reported in \protect\cite{Mowla_2019_b} the stellar mass estimates from \protect\cite{LaigleCOSMOS2016} are systematically higher than those from \cite{Skelton2014} by $\gtrsim\SI{0.1}{dex}$ for sources with $\lmass>10.75$.}.

\subsection{Sample Characterisation and Representativeness}
\label{sec:representativeness}
For sources at $z\approx 3$ the observed $H$ band probes the galaxy \ac{sed} at $\approx\SI{4000}{\angstrom}$ restframe, where the \ac{mtl} ratio is sensitive to the age of the stellar population. Selecting $H$ band bright sources as discussed in Section~\ref{sec:sample_selection} may therefore bias the sample towards younger and/or less dust attenuated stellar populations. Depending on quenching mechanisms, and at least at lower redshifts on progenitor bias effects, sizes of younger vs. older quiescent sources at fixed stellar mass may differ on average \citep[e.g.,][]{Saracco_2009, Belli_2015, Yano_2016, Williams_2017, Zahid_2017, Almaini_2017, Wu_2018}. Our $H<22$ selection could thus potentially result in a bias on the average quiescent galaxy size at a given mass inferred from this sample. In this Section we thus discuss the representativeness of this $H$-selected sample with respect to the parent (mass-selected) sample of massive quiescent galaxies at this redshift.

To address the relevance of the potential bias in the quiescent sample caused by the $H$ band selection, we compare the UVJ restframe colors of the $H<22$ quiescent population with those of the full massive galaxy population
at $2.5<z<3$.
For the full parent sample, we match the \cite{LaigleCOSMOS2016} and \cite{Muzzin_2013} catalogs in order to fit the \cite{LaigleCOSMOS2016} photometry assuming the \cite{Muzzin_2013} photometric redshifts and spectroscopic redshifts from \cite{Deugenio_2020} for our targets. We choose this approach in order to make use of the deeper photometry in the \cite{LaigleCOSMOS2016} catalog (that we use throughout in the analysis of our target sample in Section~\ref{sec:sedfitting}) and at the same time of the more accurate \cite{Muzzin_2013} photometric redshifts for massive quiescent sources at this redshift, as inferred by comparison with spectroscopic samples as shown in Appendix~\ref{sec:specz_photoz}. We use the same FAST++ setup as in Section~\ref{sec:sedfitting} to estimate stellar masses and EAZY to estimate restframe UVJ colors. We consider galaxies more massive than the mass completeness limit of $\lmass=11.1$ at $z=3$ from \cite{Muzzin_2013}, providing a sample of 43 UVJ quiescent galaxies at $2.5<z<3.0$.

\begin{figure}
	\includegraphics[width=\columnwidth]{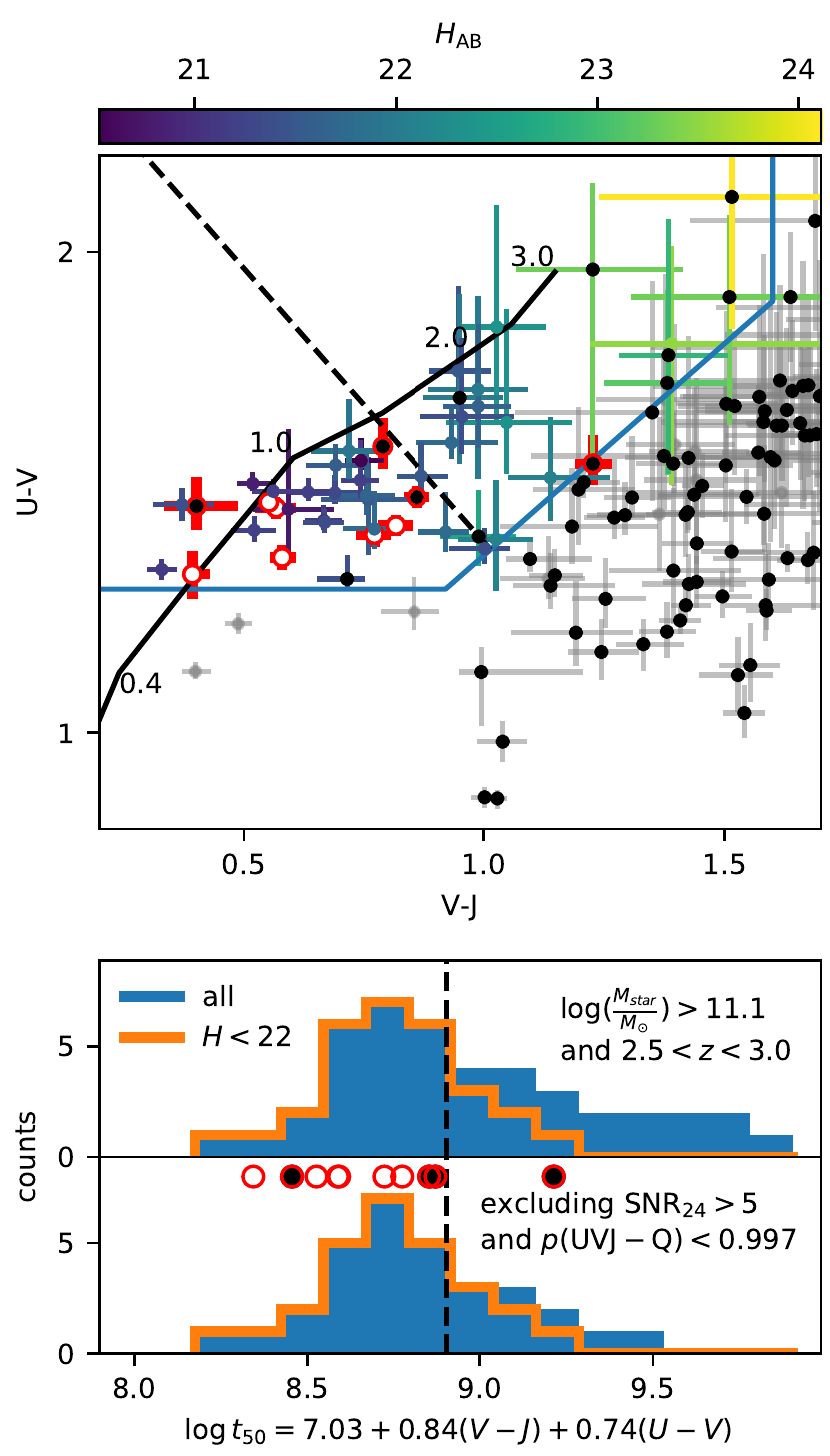}
	\caption{Top panel: restframe UVJ colors of massive ($\lmass>11.1$) galaxies in the UltraVISTA COSMOS field (see Section~\ref{sec:representativeness}) in the redshift range $2.5<z<3$. For UVJ-quiescent sources, the colors of the symbols scale with the $H$ band magnitude, as indicated by the color bar. The sources studied in this work are overplotted as red circles, with colors computed assuming the spectroscopic redshift. The 1 sigma color uncertainties account for photometric and redshift uncertainties. $\SI{24}{\micro\metre}$ detections from \protect\cite{Jin_2018} are marked with a black dot. The blue line shows the adopted separation between star-forming and quiescent galaxies \protect\citep{Williams_2009}. The black solid line shows for reference the evolution of a simple stellar population (numbers along the line show ages in Gyr).
	 The dashed line shows the location in the diagram corresponding to an average stellar age of $t_{50}=\SI{800}{Myr}$ based on the empirical relation between restframe UVJ colors and $t_{50}$ from \protect\cite{Belli_2019}. Bottom panel: histograms of the UVJ color combination translating into $ t_{50}$ with the \protect\cite{Belli_2019} relation (see Section~\ref{sec:representativeness}), for the full $\lmass>11.1$ UVJ-quiescent sample (top) and for the UVJ-quiescent sample excluding $\SI{24}{\micro\metre}$ detections with a probability of being quiescent <0.997 (see text). In both cases orange histograms refer to the $H<22$ subsample. Red symbols show the ten targets of this work.}
	\label{fig:UVJ}
\end{figure}

The location of quiescent galaxies in the UVJ plane correlates with the age of their stellar populations \citep[e.g.,][]{Belli_2019}. We thus investigate in Figure~\ref{fig:UVJ} the distribution of our targets in the UVJ plane, and more generally of sources brighter than $H=22$,  with respect to the parent population, to constrain possible biases in our sample. All of our targets are well within the UVJ-quiescent region \citep{Williams_2009} except ID 10, which is anyway consistent with being UVJ-quiescent.

\cite{Belli_2019} parametrise the relation between stellar population age and UVJ colors by adopting $t_{50}$ as an age estimate. We use this parametrization to investigate the impact of the $H$ band selection on the fraction of "post-starburst" ($t_{50}<\SI{800}{Myr}$) to old passive galaxies in the full sample of $\lmass>11.1$ quiescent galaxies. As it may be expected given the high redshift, a large fraction of the quiescent galaxy sample is made of relatively young sources which at lower redshift are typically classified as post-starburst based on their colors \citep[see also e.g.,][]{Whitaker_2012, Marchesini_2014, Merlin_2018, Maltby_2018}. To account for the uncertainties on the photometric measurements and redshift estimates we perturbe the source photometry and photometric redshift within the uncertainties and estimate restframe UVJ colors accordingly for 10000 realizations. The inferred median distribution of UVJ color combination which translates to $t_{50}$ in the \cite{Belli_2019} parametrization is shown in the bottom panel of Figure~\ref{fig:UVJ}. With this approach the estimated fraction of post-starburst galaxies in the full massive quiescent sample is $\SI{50\pm 9}{percent}$. Considering only galaxies with $H<22$ this fraction increases to $\SI{77\pm 9}{percent}$, consistent with our sample in which 9 of 10 galaxies have $t_{50}<\SI{800}{Myr}$, according to the relation from \cite{Belli_2019}.
Indeed \cite{Deugenio_2020} found $t_{50}\leq \SI{800}{Myr}$ for all galaxies in the sample. Therefore, at face value the average stellar age of galaxies in the $\lmass>11.1$, $H<22$ sample is indeed younger than in the whole $\lmass>11.1$ sample, suggesting that our sample may be more representative of younger, post-starburst quiescent systems (see Figure \ref{fig:UVJ}), and likely biased against the oldest quiescent galaxies at this redshift. If significant morphological transformations happen on longer time scales than the typical age of this sample we would not be able to see it in our analysis because our sample does not contain these older sources.
On the other hand, we stress that the uncertainties on the estimated restframe UVJ colors are significantly higher - as expected given the quality of the available photometry - for older quiescent galaxies, possibly resulting in a more significant contamination from dusty star-forming sources. To investigate this further, we also highlight in Figure~\ref{fig:UVJ} sources that are detected at $\SI{24}{\micro\metre}$ ($\approx\SI{6}{\micro\metre}$ restframe) with a $\ac{snr}>5$ in the \cite{Jin_2018} catalog. The fraction of $\SI{24}{\micro\metre}$ detected sources is higher for old quiescent galaxies than for "post-starburst" systems. Out of the six oldest UVJ quiescent sources in the sample shown in Figure~\ref{fig:UVJ}, five are detected at $\SI{24}{\micro\metre}$. However the $\SI{24}{\micro\metre}$ emission could also originate from nuclear activity (see D'Eugenio et al. in preparation for a discussion of $\SI{24}{\micro\metre}$ emissions of our targets), considering the large photometric uncertainties for the oldest sources, this suggests that a significant fraction of the full massive sample considered in Figure~\ref{fig:UVJ} might be star-forming contaminants. We thus re-estimate the fraction of post-starburst galaxies excluding all galaxies that are both $\SI{24}{\micro\metre}$ detected and UVJ quiescent with a probability lower than $p(\textrm{UVJ-Q})=0.997$ ($3\sigma$) and find that $\SI{65\pm10}{percent}$ of the full $\lmass>11.1$ sample have $t_{50}<\SI{800}{Myr}$ compared to $\SI{77\pm10}{percent}$ of the galaxies of the full sample with $H<22$.
Although some of the $\SI{24}{\micro\metre}$ detections could be due to nuclear activity, the distribution of $\SI{24}{\micro\metre}$ detections across the UVJ plane and the estimated uncertainties in UVJ colors strongly suggest that a possibly significant fraction of the oldest quiescent galaxies are actually contaminants, and that the impact of the $H<22$ selection on the age distribution of our target sample is smaller than would be suggested by face-value comparison of UVJ colors alone.
Indeed, the independent estimate of the selection bias for this sample presented in \cite{Deugenio_2020} consistently concludes that our target sample is representative of $\gtrsim\SI{70}{percent}$ of the overall quiescent population in the probed mass and redshift range. A specific - and currently very expensive - follow-up of a sample of the highest M/L ratio candidates would be necessary to conclusively address the picture of the potentially oldest massive quiescent galaxies at this redshift.

\section{Morphological analysis}
\label{sec:morphology_fit}
\newcommand{\dhst}{3D-HST}
\newcommand{\ttim}{TinyTim}

We investigate morphological properties of our targets by means of parametric modeling of the surface brightness distribution in the F160W band images. For each target we have 3 to 5 dithered observations with total exposure times ranging from $\SI{980}{s}$ to $\SI{1130}{s}$ at an observed wavelength of $\approx\SI{16000}{\angstrom}$. We reduce the preprocessed flat-fielded single exposures retrieved from the STSci archive in 2 different ways to investigate the robustness and sensitivity of the fit results to the reduction procedure. For the first reduction we use DrizzlePac release 2.2.6\footnote{https://github.com/spacetelescope/drizzlepac/blob/master/doc/source/index.rst} to subtract the background, remove cosmic rays and for each source use a square kernel to drizzle the exposures to a pixel scale of $\SI{0.06}{arcsec}$ before median stacking them to the final image. Each image covers an area of $\approx\SI{4.8}{arcmin}^2$.  For reference, the estimated $\SI{90}{percent}$ point-source completeness of the images is $\approx\SI{26.7}{mag}$.

We use SExtractor \citep{sextractor} to detect sources in the F160W band images. We select point-like sources in each image by means of a magnitude (MAG\_AUTO) vs. half-light radius (FLUX\_RADIUS $\SI{50}{percent}$) diagram.
The comparison of point-like sources across the images of the 10 different fields and at different positions on the detector suggests that the \ac{psf} is relatively stable with no significant variations for the purposes relevant to this work. This allows us to create a single \ac{psf} by stacking high \ac{snr} ($H\lesssim 21$) point-like sources from all ten fields with SWarp \citep{swarp}, improving the \ac{snr} of the model.
To estimate the effect of possible systematics of the \ac{psf} modeling on our results (see discussion in Section~\ref{sec:galfit_uncertainty}), we vary the point-like source selection criteria to create a set of \acp{psf} from our observations. We also compare these \acp{psf} with a synthetic model, created with TinyTim, and with the hybrid model from \cite{VanDerWel2014}. Both are more peaked than our models and, if fitted to point sources in our images, subtract systematically too much flux in the center, while our \ac{psf} models do not cause systematic features in the residuals, confirming that they are appropriate descriptions of the \ac{psf} of our images. A possible reason why our \ac{psf} is less sharp is the low number (3-5) of dithered exposures per target.

\begin{figure*}
	\includegraphics[height=0.98\textheight]{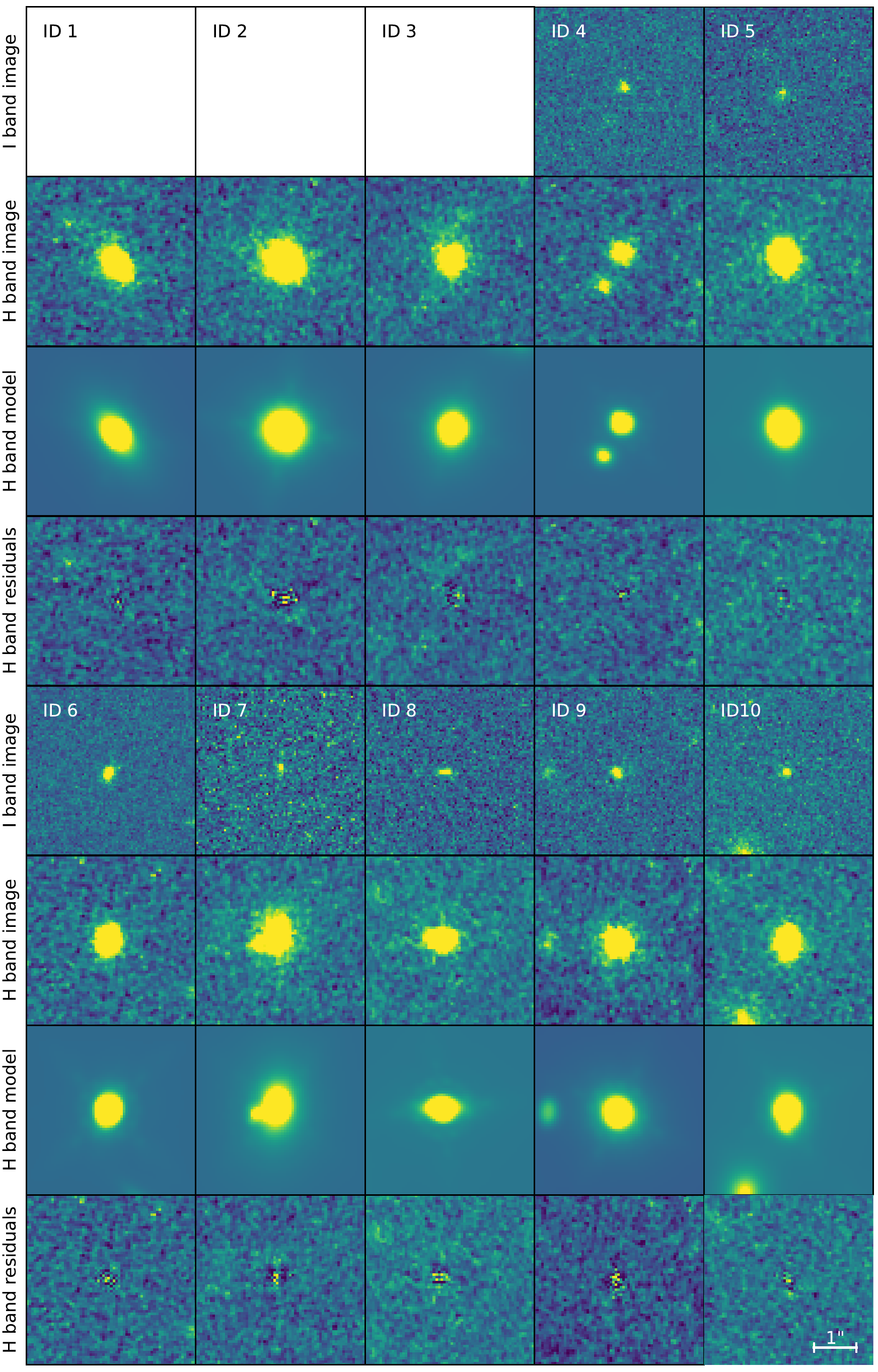}
	\caption{\ac{hst} F160W (H band) images of the ten observed targets, the best fit models (see Section~\ref{sec:morphology_fit}) and the corresponding residuals. For IDs 4-10 F814W (I~band) imaging is available and also shown. The cutouts have a size of $\SI{4}{arcsec}$ ($\approx\SI{32}{kpc}$) by side.}
	\label{fig:galfits}
\end{figure*}

For the second reduction we use the grizli pipeline\footnote{https://github.com/gbrammer/grizli/} \citep{Brammer_2018} to produce science ready images from the single exposures, detect sources and create a \ac{psf} for each image. The science images produced with the grizli pipeline are slightly sharper and have less residual cosmic rays compared to the images from the former procedure. Nonetheless, the results of our analysis are largely independent of the reduction method as discussed in detail later in this section. Cutouts of all targets are shown in Figure~\ref{fig:galfits}.

We use GALFIT \citep{galfit2002, galfit2010} to fit \ac{psf} convolved \cite{Sersic_1963, Sersic_1968} profiles to the F160W band images of the sources from both reductions. We create uncertainty maps by quadratically adding Poisson source noise to the background \ac{rms}, estimated in $9\times\SI{9}{arcsec^2}$ boxes across the images. We fit sources in cutouts with a sidelength of $\SI{9}{arcsec}$, allowing GALFIT to fit a constant background simultaneously with the \sersic\ profiles. Estimating the local background and subtracting it from the image, rather than fitting it, has no significant impact on the estimated parameters for the targets considered here. For cutouts containing multiple galaxies,  we simultaneously fit \sersic\ profiles for all sources.
We do not set prior constraints on any of the fit parameters (postition, magnitude, effective radius $\re$, \sersic\ index $n$, axis ratio $q$, position angle). Starting values for the fitted parameters are estimated based on the SExtractor output except for the \sersic\ index for which we use a starting value of 1. We verified that varying the initial parameters in a reasonable range has no impact on the results for our targets. Estimated effective radii and axis ratios are stable against the use of the different reductions and corresponding \acp{psf}, being entirely consistent within the estimated uncertainties with no systematic biases. \sersic\ indices are systematically lower by $\SI{20}{percent}$ in the grizli reduction.
In the following we always refer to the measurements obtained on the grizli reductions unless otherwise stated. We stress again that all conclusions would be unchanged if referring to the other reduction, and that all results would be fully consistent except \sersic\ indices which would be on average larger, thus resulting in even stronger conclusions on the typically high \sersic\ indices of these sources as discussed in Section~\ref{sec:results}. 
The target images, best fit models and corresponding residuals, together with \ac{hst} ACS F814W imaging \citep[][only available for IDs 4-10]{Koekemoer_2007, Massey_2010}, are shown in Figure~\ref{fig:galfits}. The resulting profile parameters are presented in Table~\ref{tab:fitresults}. Radii in this paper refer to effective radii along the semi major axis.

\newcommand{\tablecaption}{
\setlength\extrarowheight{4pt}
\setlength{\tabcolsep}{5pt}
\caption{Main properties and estimated morphological parameters of the ten targets. The column ID\textsubscript{C} lists for convenience the IDs from the COSMOS \protect\cite{LaigleCOSMOS2016} catalog. The $z_{\textrm{phot}}$ column lists the photometric redshifts used for target selection, while $z_{\textrm{spec}}$ lists the spectroscopic redshifts from \protect\cite{Deugenio_2020}. The column $\lmass$ lists stellar mass estimates (see Section~\ref{sec:sedfitting} for details, including a discussion of the uncertainties). The columns mag, $r$ (sky), $n$, $q$ list the GALFIT best fit values for the \sersic\ profile; radii are effective radii along the semimajor axis. Uncertainties are estimated as explained in Section~\ref{sec:galfit_uncertainty}. The physical radii in kpc, $\re$ ($\SI{5000}{\angstrom}$), are inferred sizes at $\SI{5000}{\angstrom}$, derived as explained in Section~\ref{subsec:mass-size-relation}.}
\label{tab:fitresults}
}

\begin{table*}
\tablecaption
\begin{threeparttable}
\begin{tabular}{cccccccccccc}
\hline \hline
ID & ID\textsubscript{C} & R.A. & Dec & $z_{\textrm{phot}}$ & $z_{\textrm{spec}}$ & $\lmass$ & mag & $\re$ (sky) & $\re$ ($\SI{5000}{\angstrom}$) & $n$ & $q$ \\
 &  & (h:m:s) & (d:m:s) &  &  &  &  & (arcsec) & (kpc) &  &  \\
\hline
1 & 135730 & 10:01:39.98 & 01:29:34.49 & 2.6 & ${2.841}_{-0.018}^{+0.021}$ & 11.14 & ${21.99}_{-0.09}^{+0.10}$ & ${0.40}_{-0.06}^{+0.07}$ & ${3.07}_{-0.47}^{+0.57}$ & ${4.7}_{-0.8}^{+0.8}$ & ${0.50}_{-0.03}^{+0.03}$ \\
2 & 137182 & 10:00:57.35 & 01:29:39.46 & 2.7 & ${2.557}_{-0.005}^{+0.005}$ & 11.27 & ${21.32}_{-0.03}^{+0.02}$ & ${0.15}_{-0.01}^{+0.01}$ & ${1.18}_{-0.04}^{+0.04}$ & ${3.0}_{-0.2}^{+0.3}$ & ${0.96}_{-0.02}^{+0.02}$ \\
3 & 252568 & 09:57:48.57 & 01:39:57.82 & 2.8 & ${3.124}_{-0.003}^{+0.003}$ & 11.32 & ${21.88}_{-0.10}^{+0.10}$ & ${0.33}_{-0.05}^{+0.08}$ & ${2.37}_{-0.37}^{+0.58}$ & ${6.2}_{-1.1}^{+1.2}$ & ${0.78}_{-0.04}^{+0.04}$ \\
4 & 361413 & 10:02:00.97 & 01:50:24.28 & 3.2 & ${3.230}_{-0.006}^{+0.007}$ & 10.75 & ${23.37}_{-0.11}^{+0.08}$ & ${0.07}_{-0.01}^{+0.01}$ & ${0.46}_{-0.07}^{+0.08}$ & ${4.8}_{-1.3}^{+2.9}$ & ${0.86}_{-0.11}^{+0.11}$ \\
5 & 447058 & 09:59:11.77 & 01:58:32.96 & 2.5 & ${2.665}_{-0.007}^{+0.003}$ & 11.11 & ${22.20}_{-0.02}^{+0.02}$ & ${0.21}_{-0.01}^{+0.01}$ & ${1.63}_{-0.05}^{+0.05}$ & ${1.2}_{-0.1}^{+0.1}$ & ${0.79}_{-0.03}^{+0.03}$ \\
6 & 478302 & 09:59:01.31 & 02:01:34.15 & 2.6 & ${2.801}_{-0.002}^{+0.005}$ & 11.13 & ${22.23}_{-0.02}^{+0.02}$ & ${0.109}_{-0.004}^{+0.004}$ & ${0.84}_{-0.03}^{+0.03}$ & ${2.3}_{-0.3}^{+0.3}$ & ${0.56}_{-0.04}^{+0.04}$ \\
7 & 503898 & 10:01:31.86 & 02:03:58.79 & 2.6 & ${2.674}_{-0.009}^{+0.005}$ & 11.32 & ${21.60}_{-0.11}^{+0.13}$ & ${0.57}_{-0.11}^{+0.13}$ & ${4.45}_{-0.89}^{+1.04}$ & ${6.3}_{-1.2}^{+1.0}$ & ${0.60}_{-0.03}^{+0.03}$ \\
8 & 575436 & 10:00:43.76 & 02:10:28.71 & 2.8 & ${2.998}_{-0.003}^{+0.002}$ & 11.17 & ${22.30}_{-0.04}^{+0.03}$ & ${0.10}_{-0.01}^{+0.01}$ & ${0.75}_{-0.04}^{+0.05}$ & ${4.3}_{-0.7}^{+0.8}$ & ${0.33}_{-0.03}^{+0.03}$ \\
9 & 707962 & 09:59:32.52 & 02:22:21.99 & 2.6 & ${2.667}_{-0.002}^{+0.015}$ & 11.3 & ${21.66}_{-0.17}^{+0.15}$ & ${0.29}_{-0.07}^{+0.16}$ & ${2.27}_{-0.58}^{+1.24}$ & 12\tnote{a} & ${0.87}_{-0.05}^{+0.05}$ \\
10 & 977680 & 10:00:12.65 & 02:47:23.47 & 2.5 & ${2.393}_{-0.000}^{+0.011}$ & 11.1 & ${22.39}_{-0.03}^{+0.03}$ & ${0.15}_{-0.01}^{+0.01}$ & ${1.19}_{-0.05}^{+0.06}$ & ${2.6}_{-0.3}^{+0.4}$ & ${0.67}_{-0.04}^{+0.04}$ \\
\hline
\end{tabular}
    \begin{tablenotes}
        \item[a] If the \sersic\ index is fixed to 4, the effective radius decreases to $\SI{0.14}{arcsec}$ ($2.1\sigma$), see Section~\ref{sec:morphology_fit}.
    \end{tablenotes}
\end{threeparttable}
\end{table*}

Some of our targets have faint close neighbouring sources with unknown redshifts \citep[see also][]{Stockmann_2020}, in particular ID 7 has a very close neighbour ($d\approx\SI{0.8}{arcsec}$, corresponding to $\SI{6}{kpc}$ if at the target redshift). With the available data, we do not see evidence of interaction between these sources.
Furthermore, after subtracting the best fit model the residuals of a single \sersic\ profile fit for IDs 2 and 10 also show faint neighbours close to the targets ($\SI{0.56}{arcsec}$ and $\SI{0.1}{arcsec}$, respectively, corresponding to $\SI{4.6}{kpc}$ and $\SI{1.0}{kpc}$ if they are at the targets redshift). For these sources an additional \sersic\ component is thus used in the following to simultaneously model the faint neighbours. The models and residuals in Figure~\ref{fig:galfits} show the modeling accounting for these sources. Estimated parameter values ($\re, n, q$) change by $\lesssim \SI{10}{percent}$, except for the estimated effective radius of ID 10 which decreases by $\SI{19}{percent}$ and its axis ratio which increases by $\SI{25}{percent}$.

For several sources, namely IDs 6, 7, 8, 9 and 10 the residual images show a central residual. For ID 8 about $\SI{4}{percent}$ of the pixels associated with the source have a significance of more than $3\sigma$ in the residuals. For all other sources this fraction is $\lesssim\SI{2}{percent}$. This is also shown in the Appendix in Figure~\ref{fig:residuals_significance}. We verified that this is not a \ac{psf} effect by fitting the targets with different \ac{psf} models, including the more peaked TinyTim and \cite{VanDerWel2014} models and examining a larger number of residuals from fits of other sources in the images. We also verified that adding an additional point-like component to the fit does not produce an appreciable improvement on the residuals for any of these targets.
We note that 3 of these sources (IDs 6, 7 and 10) are detected in both the \cite{Jin_2018} $\SI{24}{\micro\metre}$ catalog and the \cite{Marchesi_2016} catalog of X-ray sources. Therefore, some level of star formation (in the galaxy center) and/or nuclear activity might possibly cause the central excess. By comparing the $\SI{24}{\micro\metre}$ and X-ray luminosities, D'Eugenio et al. (in preparation) suggest that in at least two out of these three sources the IR and X-ray emission is likely \ac{agn} dominated.
The flux of the central residual for these sources is smaller than $\SI{10}{percent}$ of the image flux in any pixel.

\subsection{Uncertainties}
\label{sec:galfit_uncertainty}
The precision and accuracy of our measurements are limited by noise and by uncertainties in the \ac{psf} model, which are analysed in the following.

To estimate the impact of \ac{psf} uncertainties on the parameter estimates we create on empty images without noise artificial sources with effective radius ranging from $0.03$ to $\SI{0.72}{arcsec}$ ($\SI{0.2}{kpc}\leq \re \leq\SI{5.8}{kpc}$ at $z=2.75$) and with \sersic\ indices from $0.5$ to $8$, spanning a reasonably wide range of \sersic\ parameters for quiescent galaxies in the mass and redshift range of our targets. We then fit these artificial sources with the same procedure used for our targets, using different \ac{psf} models (see Section~\ref{sec:morphology_fit}) for the convolution in the creation and in the fitting process.
The deviations of the retrieved parameters (effective radius, \sersic\ index, axis ratio) from the input ones are generally smaller than $\SI{5}{percent}$; for very small radii ($\re\lesssim\SI{0.05}{arcsec}$ or $\re\lesssim\SI{0.4}{kpc}$ in the probed redshift range) they can exceed $\SI{10}{percent}$. All of our targets except one (ID 4, $\re\approx\SI{0.07}{arcsec}$) are significantly larger than this size. The uncertainties on the \ac{psf} model are therefore expected to have a subdominant impact on our results.

To investigate the uncertainties due to noise we create \sersic\ models with Poisson noise in different empty areas of the observed images. We first create sources with the approximate magnitude of our targets, $H\approx 22$, and effective radius and \sersic\ index in the same range as discussed above for the evaluation of systematic uncertainties.
 In this case we use the same \ac{psf} for the creation of the artificial sources and for their modeling.

\begin{figure*}
	\includegraphics[width=\textwidth]{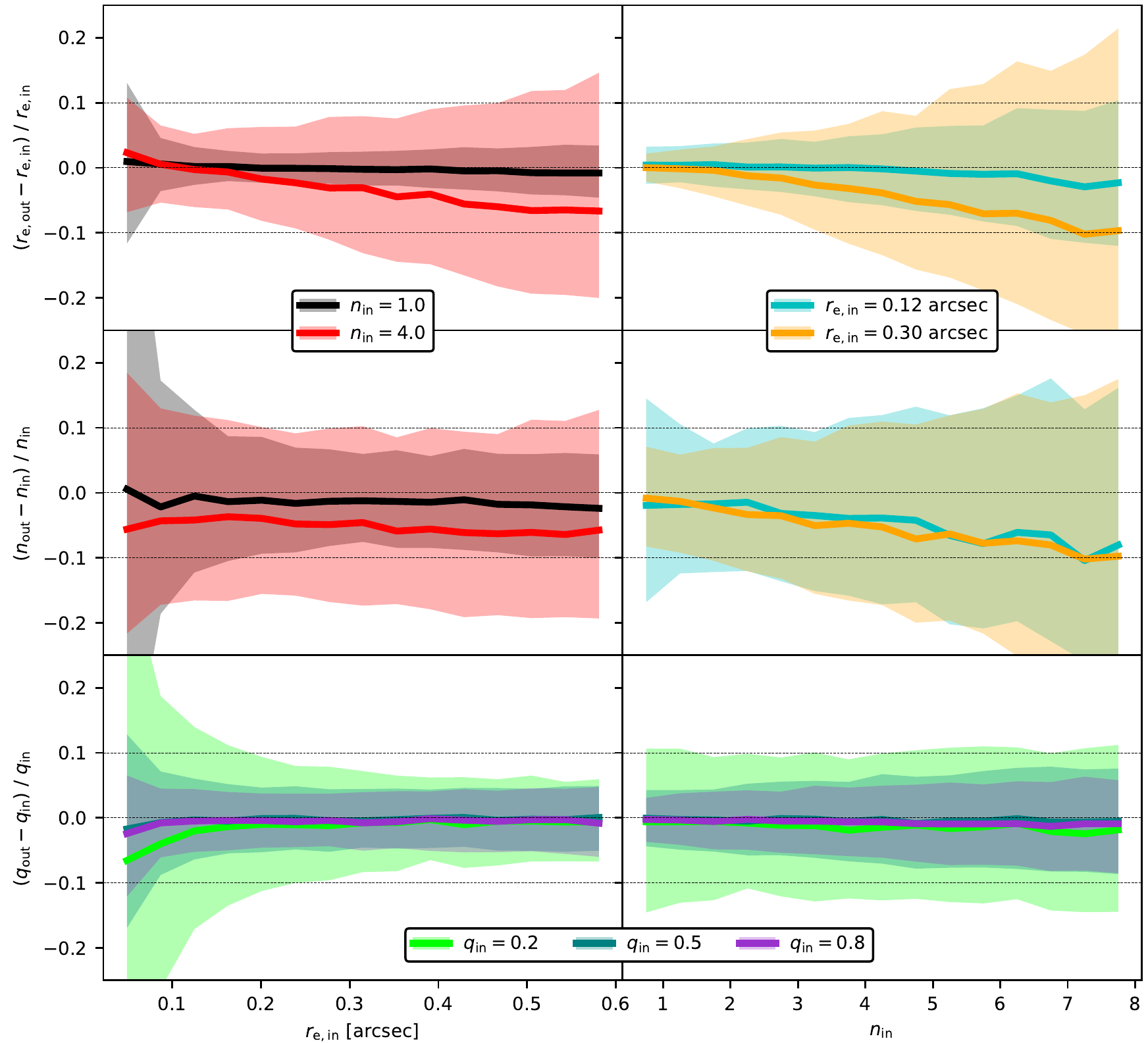}
	\caption{Relative uncertainties on the estimated effective radius, \sersic\ index and axis ratio from the simulations (Section~\ref{sec:galfit_uncertainty}), defined as $(p_{\textrm{out}}-p_{\textrm{in}})/p_{\textrm{in}}$ where $p_{\textrm{in}}$ is the input parameter value of the artificial source and $p_{\textrm{out}}$ the retrieved value from the fit. Left and right-hand panels show, respectively, the deviation of the retrieved vs. input parameters as a function of effective radius and \sersic\ index of the source. The solid lines represent the median deviation from the input value, the shaded areas show the scatter estimated from the $16-84$ percentile range. We show as an example in the upper two panels results for a \sersic\ index of 1 (black) and 4 (red) in the left panels and for $\re=\SI{0.06}{arcsec}$ (cyan) and $\re=\SI{0.12}{arcsec}$ (orange) in the right panels, both with axis ratios $0.5\leq q_{in}\leq1$. In the lower panels we show uncertainties on the retrieved axis ratio for input axis ratios of $0.2,0.5$ and  $0.8$, as indicated.}
	\label{fig:simulation_uncertainties}
\end{figure*}

In Figure~\ref{fig:simulation_uncertainties} we show the deviation of the retrieved best-fit values for the \sersic\ index, effective radius and axis ratio with respect to the input values. The estimated scatter in $\re$ and $n$ is of the order of $\SI{10}{percent}$ but depends on the actual values of $\re$ and $n$. The scatter of the axis ratio is of the order of $\SI{5}{percent}$ except for very small ratios of $\approx 0.2$ where it reaches $\approx\SI{10}{percent}$. Models with larger \sersic\ indices have generally larger uncertainties on the retrieved parameters and the \sersic\ index tends to be underestimated, because of the extended tails and low \ac{snr} in the outskirts \citep[see also e.g.,][]{Marleau_1998,Pignatelli_2006, Sargent_2007, Pannella_2009}. This underestimation, $\Delta n \approx \SI{5}{percent}$ for $n\approx 4$ and $\Delta n \approx\SI{10}{percent}$ for very high \sersic\ indices of $n\approx 8$, is about 4 times smaller than the statistical uncertainty. Sources with very small radii, close to the resolution limit, are affected by larger uncertainties, as well as those with very large radii because of the lower \ac{snr} per pixel at fixed magnitude. Sources with $n\approx 1$ and with $\re\gtrsim \SI{0.15}{arcsec}$ have uncertainties $\sigma_{\re}<\SI{5}{percent}$ and $\sigma_n < \SI{10}{percent}$. The effective radius of models with large $n$ and $\re$ is also affected by a systematic underestimation of $\re$. For large sources with $\re=\SI{0.3}{arcsec}$ and $n\approx 4$, $\re$ is underestimated by $\approx \SI{5}{percent}$, for $n=8$ by $\approx \SI{10}{percent}$. These systematics are small compared to the statistical uncertainties for the same models.

To properly estimate statistical uncertainties on the measured \sersic\ parameters for each target, we then create artificial sources with parameters in a $\SI{10}{percent}$ range around the best fit models, motivated by the previous results, adding as usual Poisson noise. We add them to different empty areas of the observed images and fit them again with the same \ac{psf} as used for the creation. Since the estimated systematics are typically small compared to the statistical uncertainties we do not apply any correction for the described systematics in the following.

\section{Results and discussion}
\label{sec:results}

\subsection{Broad structural properties}
\sersic\ indices of our sources range from 1.2 to 6.3 with median statistical uncertainties of $\SI{16}{\percent}$ except for source ID 9, which formally has a best-fit \sersic\ index of 12. The median \sersic\ index of all targets is $\asunc{4.5}{1.4}{0.3}$. The high \sersic\ index of ID 9 has no strong influence on the median \sersic\ index, excluding it leads to a median of $n=4.3_{-1.2}^{+0.5}$.
The low \sersic\ index $n=\asunc{1.2}{0.1}{0.1}$ of ID 5 is also reinforced by the diffuse appearance in the F814W image compared to the other sources (see Figure~\ref{fig:galfits}).

\begin{figure*}
	 \includegraphics[width=\textwidth]{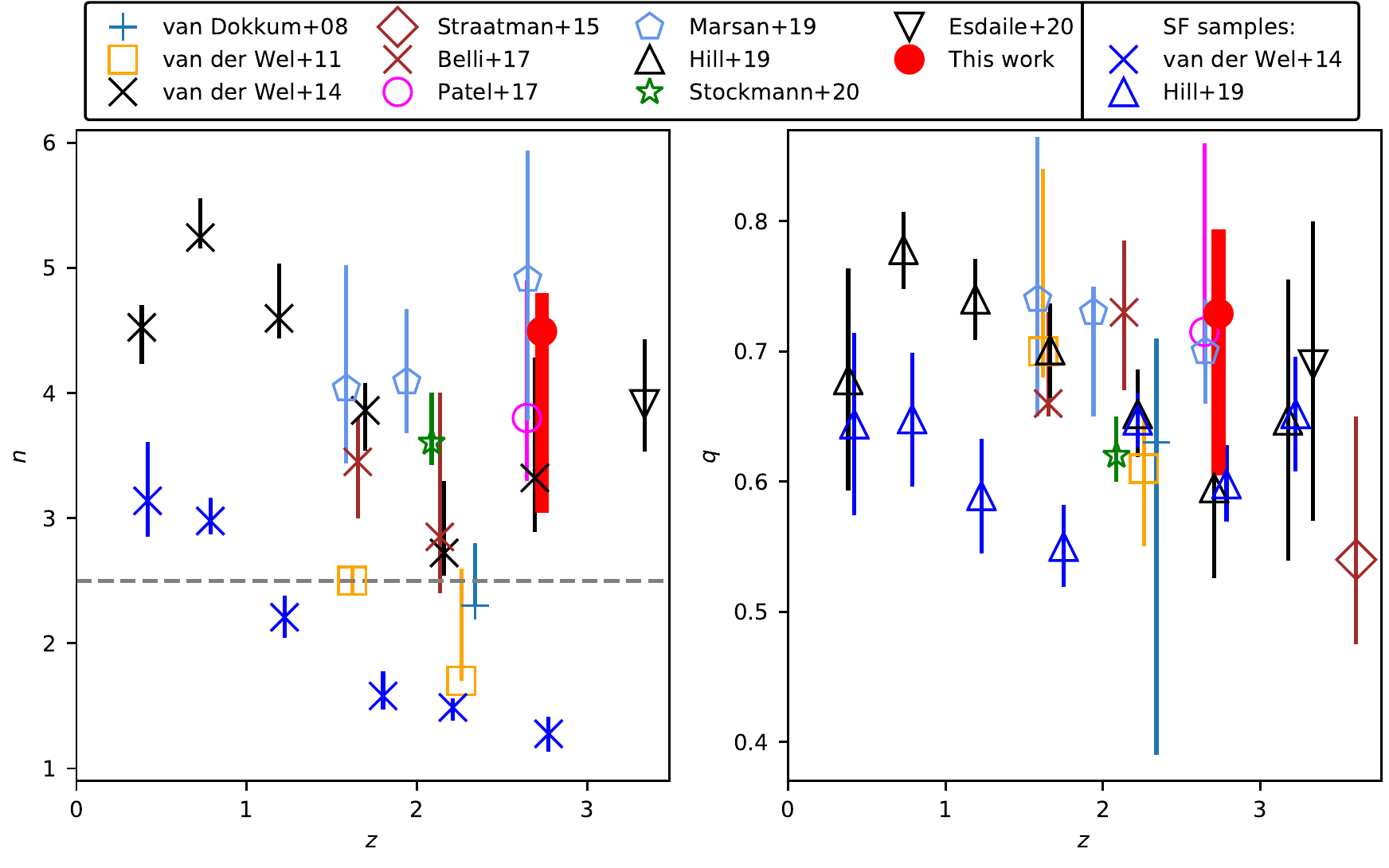}
	 \caption{Median \sersic\ indices (left panel) and axis ratios (right panel) of massive quiescent galaxy samples as a function of redshift. For comparison, we also show the evolution of the \sersic\ index for star-forming galaxies from \protect\citet[][left]{VanDerWel2014} and of the axis ratios from the same sample analysed by \protect\citet[][right]{Hill_2019}. Sources from \protect\cite{VanDerWel2014} have stellar masses of $11<\lmass <11.5$ and sources from \protect\cite{Hill_2019} have $\lmass>11$. Results from other works are reported as published, with no further mass selection applied. If the same mass selection is applied to the other samples, measurements change within the uncertainties with no systematics. Uncertainties for the samples of \protect\cite{VanDerWel2014} and \protect\cite{Hill_2019} are taken from the papers while we use bootstrapping for the calculation for the other samples.
	 }
	 \label{fig:nevolution}
 \end{figure*}

In Figure~\ref{fig:nevolution} we show the median \sersic\ index of massive quiescent galaxy samples as a function of redshift. For comparison to lower redshift quiescent and star forming galaxies we show median \sersic\ indices of galaxies with stellar masses of $11<\lmass<11.5$ from the morphological analysis of \cite{VanDerWel2014}. At all redshifts, median \sersic\ indices of quiescent galaxies are significantly larger than those of star-forming galaxies. The median \sersic\ indices of quiescent galaxies from the \cite{VanDerWel2014} sample decrease from $n=\asunc{4.5}{0.3}{0.2}$ at $z=0.4$ to $n=\asunc{3.3}{0.4}{1.0}$ at $z=2.7$.
Our results are consistent with no significant evolution in the median \sersic\ index of quiescent galaxies up to $z\approx 3$, in agreement with other studies from \cite{Patel_2017}, \cite{Mowla_2019_b}, \cite{Marsan_2019}, \cite{Stockmann_2020} and \cite{Esdaile_2020}, although some investigations have reported lower \sersic\ indices at $z\gtrsim 1.5$ \citep[e.g.,][]{VanDokkum_2008, VanDerWel_2011}.

Axis ratios of our targets range between 0.33 and 0.96 with a median of $\asunc{0.73}{0.12}{0.06}$. The only source with $q<0.5$ is ID 8 having $q=\asunc{0.33}{0.03}{0.03}$ in spite of a high \sersic\ index of $\asunc{4.3}{0.7}{0.8}$. The source also appears rather flat in the F814W image, which could be explained by a combination of an older bulge with redder colors and a younger and bluer disc that is seen edge-on.

\cite{Hill_2019} find a redshift and stellar mass independent linear relation between \sersic\ index and apparent axis ratio with a slope of $\textrm{d}q / \textrm{d}n = 0.062$, yielding an axis ratio of $0.71$ at our median \sersic\ index of $4.5$, in perfect agreement with our measurement and reinforcing our conclusions on the generally high \sersic\ indices of these sources.

In Figure~\ref{fig:nevolution} we also show the median axis ratio of massive quiescent galaxies as a function of redshift from \cite{Hill_2019}, comparing with measurements from our work and other studies at $z\gtrsim 1.5$.
While at $z<2$ median axis ratios of quiescent galaxy samples are larger than those of star-forming galaxies, at $z\gtrsim 2$ no clear difference can be seen, although uncertainties become large and quiescent galaxy sample contamination from starforming sources is likely more significant.
Our measurement of the average axis ratio is in agreement with typical axis ratios of quiescent galaxies at low redshift \citep[e.g.,][]{Holden_2012, Hill_2019}. Consistent with our measurements, \cite{Patel_2017} and \cite{Marsan_2019} also find high axis ratios and \sersic\ indices for massive ($\lmass>11.26$) quiescent galaxies at $z\approx 2.6$ as well as \cite{Esdaile_2020} at $z\approx 3.3$, suggesting that already at $z\approx 3$ a large fraction of quiescent galaxies are bulge dominated.
On the other hand \cite{VanDokkum_2008} investigated morphologies of 9 spectroscopically confirmed massive ($\lmass > 11.1$) quiescent galaxies at $z\approx 2.3$. The median \sersic\ index of their sample is $\asunc{2.3}{0.0}{0.5}$ and the median apparent axis ratio $\asunc{0.63}{0.24}{0.08}$.
In agreement with these results, \cite{VanDerWel_2011} analysed a color selected sample of 14 massive ($\lmass > 10.8$) quiescent galaxies at $1.5 < z < 2.5$ finding a median \sersic\ index of $\asunc{2.45}{0.40}{0.15}$ and a median axis ratio of $\asunc{0.67}{0.06}{0.10}$. \cite{Belli_2017} find a median \sersic\ index of $\asunc{3.25}{0.30}{0.45}$ and a median axis ratio of $\asunc{0.69}{0.04}{0.05}$ in the same redshift range. \cite{Hill_2019} investigated the median flattening of galaxies in the redshift range $0.2<z<4.0$, based on the structural analysis from \cite{VanDerWel2014} and also find that, for quiescent galaxies with $\lmass>11.0$, the apparent axis ratio decreases to $q=0.60\pm0.07$ at $z=2.7$. In contrast to results from this work and other previous investigations as discussed above, these studies suggest that massive quiescent galaxies at high redshift are flatter than low-redshift counterparts, with a large fraction of disk-dominated systems.
Considering at face value our results on both \sersic\ indices and axis ratios, our measurements do not lend support to this picture. Nonetheless, concerning axis ratios, we note that given the large statistical uncertainties our median axis ratio is still consistent with results from \cite{VanDokkum_2008, VanDerWel_2011, Hill_2019}.
 Furthermore, one of our targets has $q<0.5$ and three have $n<3$, suggesting that some sources in our sample might indeed be disk-dominated or have a significant disk component.
We note that differences in results and conclusions from the studies discussed above may partly derive from different sample selection criteria (in particular \cite{VanDerWel_2011} and \cite{Hill_2019} rely on different flavours of photometrically selected samples).

\subsection{The mass-size relation}
\label{subsec:mass-size-relation}
The estimated effective radii of the galaxies in our sample are between $0.07$ and $\SI{0.57}{arcsec}$, corresponding to physical sizes of $0.5$ to $\SI{4.5}{kpc}$ at restframe wavelengths from $3800$ to $\SI{4700}{\angstrom}$.
Fitting ID 9 with the \sersic\ index fixed to a typical value for bulge dominated systems of $n=4$ (close to the sample median of $n=\asunc{4.5}{1.4}{0.3}$) leads to a decrease of the estimated effective radius by $\SI{50}{percent}$.

 Because of - mostly negative - color gradients of galaxies  \citep{Szomoru_2011, Wuyts_2012, VanDerWel2014, Suess_2019, Suess_2019b}, galaxy sizes inferred from light profiles depend on the probed wavelength with sizes being larger at shorter wavelengths. For a proper comparison with previous works we convert all measured sizes to the same restframe wavelength of $\SI{5000}{\angstrom}$, adopting the correction appropriate for quiescent galaxies from \cite{VanDerWel2014}:
\newcommand{\unit}[1]{\,\mathrm{#1}}
\begin{equation}
\re(\SI{5000}{\angstrom}) = \re(\lobs) \left(\frac{1+z}{\lobs/\SI{5000}{\angstrom}}\right)
^{\dlrdll}
\end{equation}
with
\begin{equation}
\dlrdll = -0.35 + 0.12z - 0.25\log\left(\frac{\Mstar}{10^{10}\Msol}\right),
\end{equation}
where $\lobs$ is the observed wavelength of $\SI{16000}{\angstrom}$. This results in a very small correction decreasing the measured sizes of our targets by about $\SI{5}{percent}$; the final sizes adopted in the following are between $0.5$ and $\SI{4.4}{kpc}$ with a median size of $\asuncunit{1.4}{0.2}{0.9}{kpc}$. Uncertainties on the median are obtained by bootstrapping. These sizes are reported in Table~\ref{tab:fitresults}.

\begin{figure}
	\includegraphics[width=\columnwidth]{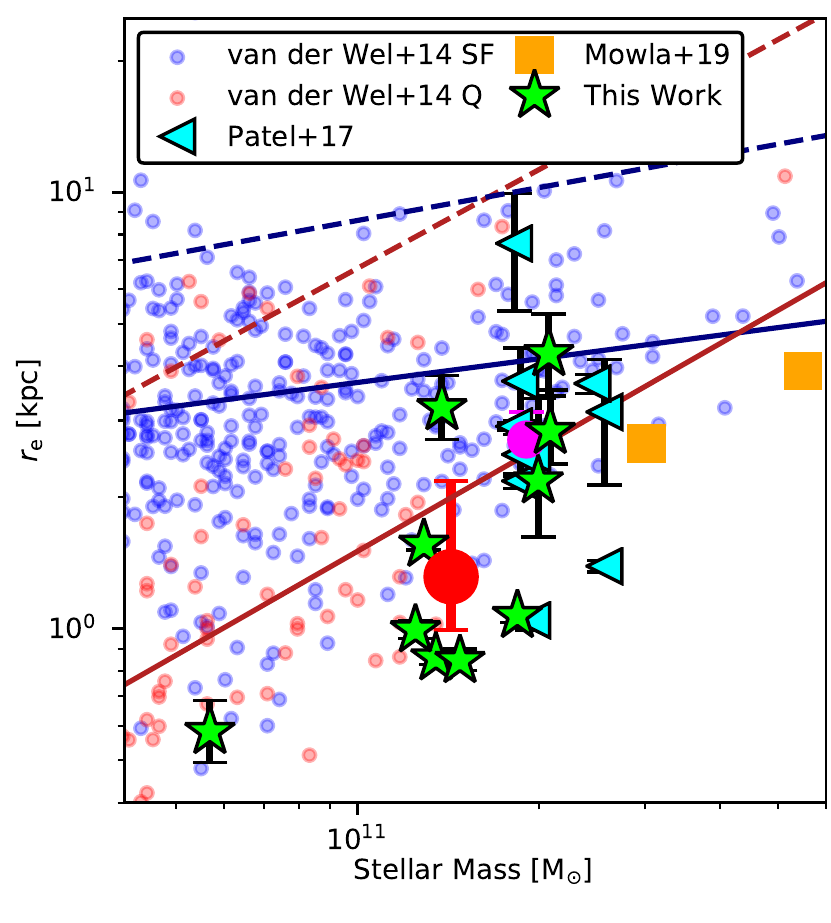}
	\caption{
	Estimated effective radii as a function of stellar mass. Sizes of individual galaxies from this work (stars),  \protect\citet[][triangles]{Patel_2017} and \protect\citet[][squares]{Mowla_2019_b} are scaled to a pivot redshift of $2.75$ (see Section~\ref{subsec:mass-size-relation}). The red and purple dots show the median size and median mass from our sample and from \protect\cite{Patel_2017}, respectively. Red (blue) dots are quiescent (star-forming) galaxies with $2.5<z<3.0$ from \protect\cite{VanDerWel2014}, with red and blue solid lines showing the corresponding best-fit mass-size relations. For comparison, the best-fit relations for $z<0.5$ from the same work are shown with dashed lines.}
	\label{fig:M-R}
\end{figure}
In Figure~\ref{fig:M-R} we compare our results for the stellar mass vs. size relation of quiescent galaxies at $z\approx 3$ with previous measurements of photometrically (UVJ) selected quiescent and star-forming galaxies from \cite{VanDerWel2014}, as well as quiescent galaxies from \cite{Patel_2017} and \cite{Mowla_2019_b}, in the same redshift range.
To ensure that no systematics on stellar masses affect our comparison with size estimates from \cite{VanDerWel2014}, we fit \sersic\ profiles to all galaxies in our fields and estimate their stellar masses as explained in Section~\ref{sec:sedfitting}. Their mass size relations are consistent with results from \cite{VanDerWel2014} at the corresponding redshifts, indicating no significant systematics between the mass and size measurements in the two studies.
For a more proper comparison of the mass-size relation within the probed $2.4<z<3.2$ range, and given the small sample size of the plotted samples from this work, \cite{Patel_2017} and \cite{Mowla_2019_b}, we scale individual sizes for galaxies from these samples to a pivot redshift of $2.75$ using the size evolution dependence on the Hubble parameter from \cite{VanDerWel2014}. This scaling leads to a maximum decrease of sizes by $\SI{17}{percent}$ at the lowest redshift $z=2.4$ and a maximum increase by $\SI{25}{percent}$ at the highest redshift $z=3.2$ \footnote{If rather than using the size evolution dependence on the Hubble parameter we use the dependence on $1+z$, always from \cite{VanDerWel2014}, the maximum increase (decrease) is $\SI{14}{percent}$ ($\SI{20}{percent}$) which does not impact the results of this analysis.}. The median sizes of our, \cite{Patel_2017} and \cite{Mowla_2019_b} samples decrease by $\approx 7$, $8$ and $\SI{10}{percent}$, respectively.

Our sources specifically probe the mass-size relation at the highest stellar masses, for the first time with a statistical, homogeneously analysed, spectroscopically confirmed quiescent galaxy sample at this redshift. Our  measurements thus extend towards the highest masses the determination of the mass-size relation of quiescent sources at $z\approx 3$, which in deep fields is typically dominated by lower-mass galaxies because of the intrinsically low number density of very massive quiescent sources.
Our measurement of the median quiescent galaxy size at the tip of the mass-size relation ($\lmass\gtrsim 11$) at $z\approx 3$ is nonetheless consistent with the relation measured in \cite{VanDerWel2014}.

\subsubsection{Central Stellar Mass Densities}
Although with the available data we can only probe the projected surface brightness distribution of our targets in the F160W band, we attempt a conversion of the observed light profile to a stellar mass density profile, to estimate central densities of these galaxies for the purpose of comparing with other similar studies. This conversion relies in particular on the assumption that the observed F160W light traces stellar mass across the galaxy:  we stress that, also given the restframe wavelength probed by the F160W imaging at the redshift of these sources, this assumption has in fact significant limitations which are neglected in the following calculations.

We deproject the observed surface brightness distributions of our targets and calculate their central densities within $\SI{1}{kpc}$ following the procedure in \cite{Whitaker_2017}. Briefly, we calculate a circularized density profile from the best-fit structural parameters derived in Section~\ref{sec:morphology_fit} by performing an Abel transform as descibed in \cite{Bezanson_2009}. In the assumption that light traces mass the central stellar mass density within $\SI{1}{kpc}$ is then given by:
\begin{equation}
\rho_1=
\frac{
\int_{0}^{\SI{1}{kpc}} \rho(r)r^2\mathrm{d}r
}{
\int_{0}^{\infty} \rho(r)r^2\mathrm{d}r
}
\frac{
\Mstar
}{
\frac{4}{3}\uppi (\SI{1}{kpc})^3
},
\end{equation}
where $\rho(r)$ is the spherical density profile as a function of radius.
We estimate uncertainties coming from the measurement of structural parameters by perturbing $\re$, $n$ and $q$ within their estimated uncertainties and recalculating the central densities 1000 times. These uncertainties are at most $\SI{0.09}{dex}$; the uncertainties on the central densities are therefore dominated by the uncertainties on the stellar mass estimates (see Section~\ref{sec:sedfitting}) as well as by the limitations of the adopted assumptions to convert the observed surface brightness distribution to a stellar mass density profile.
The central densities of the targets are $9.8 \lesssim \lrho \lesssim10.4$ with a median of $\lrho=10.1\pm0.1$.
Such central densities translate in circular velocities at $r=\SI{1}{kpc}$  of $\SI{330}{km/s} \lesssim \vc \lesssim \SI{640}{km/s}$ (median $\asunc{480}{60}{50}{\textrm{km/s}}$), as obtained by $\vc = \sqrt{\frac{4\uppi}{3}(\SI{1}{kpc})^2\rho_1\textrm{G}}$, where G is the gravitational constant, by balancing gravitational and centrifugal forces \citep[see e.g.,][]{Whitaker_2017}.

These high inferred central densities - and implied circular velocities - are in line with previous determinations for high-redshift massive, quiescent sources \citep[e.g.,][]{Dokkum_2014, Whitaker_2017, Mowla_2019_b}.

\subsection{Size evolution}
\label{subsec:z-size-relation}
\begin{figure}
	\includegraphics[width=\columnwidth]{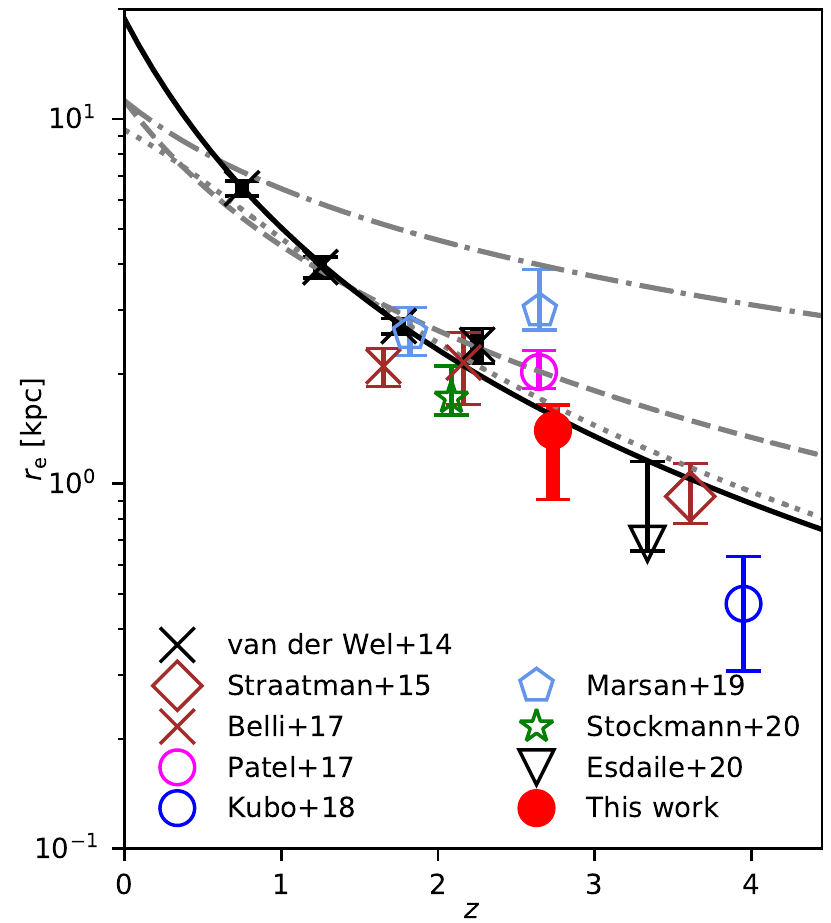}
	\caption{Median effective radii of quiescent galaxy samples with $10.6<\lmass<11.8$ as function of redshift. Sizes are scaled to a pivot mass of $\lmass=11.1$ as explained in section~\ref{subsec:mass-size-relation}. The \protect\cite{Marsan_2019} \protect\citep{Belli_2017} sample is split into two subsamples with $1.5<z<2.1$ $(1.5<z<1.9)$ and $2.6<z<3$ $(2.1<z<2.4)$. The dotted line shows the best fit model from \protect\cite{VanDerWel2014} assuming a $\re\propto h(r)^{\beta}$ relation, while the dashed and solid lines show best fit models assuming $\re\propto (1+z)^{\beta}$ from \protect\cite{VanDerWel2014} and \protect\cite{Kubo2018}, respectively. The dashed-dotted line shows for reference the median evolution for star-forming galaxies from \protect\cite{VanDerWel2014}.}
	\label{fig:redshift_evolution}
\end{figure}

To constrain the redshift evolution of massive quiescent galaxies at early times we compare sizes from our work with measurements from \cite{VanDerWel2014}, \cite{Straatman2015}, \cite{Patel_2017}, \cite{Kubo2017}, \cite{Belli_2017}, \cite{Kubo2018}, \cite{Marsan_2019}, \cite{Stockmann_2020} and \cite{Esdaile_2020} in Figure~\ref{fig:redshift_evolution}.
Sources from \cite{Belli_2017}, \cite{Stockmann_2020}, \cite{Esdaile_2020} and from this work are spectroscopically confirmed quiescent galaxies, while the other studies we compare with rely on photometrically selected quiescent sources.

Sources from \cite{VanDerWel2014} in this figure have masses between $11.0<\lmass<11.5$ with a median mass of $\lmass\approx 11.1$ in each redshift bin. We use the redshift independent slope of the mass size relation from \cite{VanDerWel2014} ($\mathrm{d}\log(\re)/\mathrm{d}\log(\Mstar)=0.7$) to scale sizes of individual galaxies of all other samples (with masses in the range $10.6 < \lmass < 11.8$) to $\lmass=11.1$. We then calculate median sizes and uncertainties for all samples by bootstrapping. The aforementioned scaling to a common mass of $\lmass=11.1$ has a very limited impact on our measurement of the median size of our sample being $\asuncunit{1.4}{0.5}{0.2}{kpc}$ at the pivot mass, basically affecting only the uncertainties.
For the least massive sample \citep{Straatman2015} this correction increases the median size by $\approx\SI{0.1}{dex}$, while for the most massive sample \citep{Marsan_2019} the median size decreases by $\approx\SI{0.25}{dex}$.
For the study of \cite{Kubo2018} the size and uncertainty of the stack is shown. All points are plotted at the median redshift of the respective sample\footnote{
We do not use any size vs. redshift relation to scale individual galaxy sizes to the median redshift before calculating the median size. However, if any of the relations shown in Figure~\ref{fig:redshift_evolution} is adopted to do so, the impact on the median size is smaller than $\SI{0.03}{dex}$ for any sample and does not affect our discussion.
 }.

Our measurements are in line with previous determinations indicating that sizes of quiescent galaxies at fixed stellar mass have increased by nearly one order of magnitude since $z=3$.
 The different models from \cite{VanDerWel2014} and \cite{Kubo2018}, that parametrise the redshift evolution of the mass-size relation either as a function of the Hubble parameter, that is related to halo properties, or as a function of the scale factor, differ by a maximum of $\SI{0.1}{dex}$ at the median target redshift of 2.73.
The median of our size measurements is consistent with the \cite{VanDerWel2014} evolution as a function of $h(z)$ and the \cite{Kubo2018} evolution as a function of $1+z$. Our measurement is 2 sigma smaller than expected from the \cite{VanDerWel2014} evolution as a function of $1+z$, which suggests, together with the higher redshift measurements by \cite{Straatman2015}, \cite{Kubo2018} and \cite{Esdaile_2020} that size evolution is steeper than in this relation.
 
At redshifts closest to our measurements, the median sizes from \cite{Patel_2017} and especially from the highest redshift sources in \cite{Marsan_2019} tend to be larger than our estimate as well as than extrapolations from most of the other high-redshift measurements discussed above.
Both measurements are largely based on the same sample of galaxies with a median mass of $\lmass\approx 11.3$.
Based on these measurements, \cite{Patel_2017} and \cite{Marsan_2019} suggest that very massive galaxies with $\lmass>11.25$ are systematically larger than expected from the mass-size relation determined at lower masses, and thus that the size evolution factor may be different at the highest masses.
On the other hand, results from the very massive samples with $\lmass\approx 11.5$ from \cite{Marsan_2019} at $z\approx 1.8$ and from \cite{Stockmann_2020} at $z\approx 2$ -- the latter likely affected by minimal contamination from star-forming sources, due to spectroscopic confirmation -- do not seem to support such a scenario.
From the four most massive galaxies in our sample with $\lmass>11.25$ we see excellent agreement with the extrapolation of the \cite{VanDerWel2014} mass-size relation, although the statistics are very limited due to the small sample size.

\section{Summary and Conclusions}
\label{sec:summary}
We have analysed structural properties of a first sizeable sample of spectroscopically confirmed, massive, quiescent galaxies at $z\approx 3$  \citep{Deugenio_2020}. Due to the rarity of these objects, we relied on targeted \ac{hst}/\ac{wfc3} imaging of 10 robust candidates.

We estimate structural properties by fitting \sersic\ profiles to the F160W images and obtain half light radii of about $\re\approx\SI{1}{kpc}$ at stellar masses of $\lmass\approx 11.2$, in agreement with photometrically selected samples at this redshift. The comparison with sizes of massive quiescent galaxies at different redshifts shows substantial agreement with the expected evolution of the mass-size relation of quiescent galaxies as determined in previous work, pointing towards a size evolution factor at fixed stellar mass of almost a factor 10 from $z\approx 3$ to today.

Although our observations are consistent with a fraction of our sample being made of disk-dominated galaxies, and a larger sample would be needed to better quantify the prevalence of such sources, our measurements of both axis ratios and \sersic\ indices suggest that massive, quiescent galaxies are already largely bulge dominated at $z\approx3$.
Based on a sample of massive galaxies in the redshift range $0.5<z<3$, \cite{Barro_2017} find a redshift and mass independent relation between the offset of a galaxy's star formation rate from the main sequence and the central mass density, that strongly correlates with \sersic\ index. This implies that star forming galaxies first grow inside out while increasing the radius and the central mass density, followed by a phase of enhanced bulge growth that increases the \sersic\ index. Star formation is then suppressed and galaxies become quiescent. This picture is also in line with other studies by e.g. \citet{Lang_2014}, \cite{Dokkum_2014, Dokkum_2015}, \cite{Gobat_2017}, \cite{Whitaker_2017} and \cite{Gomez-Guijarro_2019} and is consistent with the large fraction of bulge dominated systems in our sample. The presence of bulge-dominated, quiescent galaxies already at $z\approx 3$ constraints the timescales of quenching and of morphological transformations at early times. Although merging is believed to be a critical process to explain the size and structural evolution of quiescent high redshift progenitors into local massive ellipticals, the combination of young ages and dense, compact structures of the most distant quiescent galaxies such as those studied here suggest different mechanisms for the fast formation of the stellar core. Matching number densities and high \acp{sfr} at high redshift have suggested an evolutionary path linking the intense bursts of star formation in high-redshift sub-mm galaxies to the high stellar densities and old stellar populations of massive elliptical galaxies at lower redshifts down to the nearby universe \citep[e.g.,][]{Lilly_1999, Genzel_2003, Tacconi_2008, Cimatti_2008, Simpson_2014, Simpson_2017}, including in particular the most distant massive, dusty star-forming galaxies being likely progenitors of (at least some of) the first massive, compact quiescent galaxies at $z>2$ \citep[e.g.,][]{Toft_2014,  Valentino_2020, Forrest_2020}. High resolution ALMA imaging of z~4-6 dusty, massive, highly star-forming sources confirms the existence of possible star-forming progenitors with already compact morphologies \citep{Oteo_2017, Jin_2019}. The majority of bright sub-mm galaxies in simulations \citep{Hopkins_2008, Dekel_2009b, Zolotov_2015, Wellons_2015, Lagos_2020} are experiencing central starbursts driven by two main channels, gas-rich major mergers and disk instabilities, that increase the central density forming a compact remnant. Such remnants may still have disks and disk-dominated kinematics \citep[e.g.,][and references therein]{Belli_2017, Toft_2017, Newman_2018}, suggesting that the morphological transformations creating dispersion-supported ellipticals are not necessarily coincident with quenching. The mechanism by which star formation would stop in the compact star-forming progenitors is still unclear, with proposed processes including dynamical heating \citep["morphological quenching",][]{ Martig_2009}, stellar and \ac{agn} feedback \citep{Hopkins_2006}, shock heating \citep{Dekel_2006}, cosmological starvation \citep{Feldmann_2015}, starvation by the circumgalactic medium having too high angular momentum to be accreted by the central galaxy \citep{Peng_2020} \citep[see also e.g.,][and references therein]{Man_2018}. Some observations \citep{Nelson_2014, Gilli_2014} have identified possible compact star-forming progenitors suggestive of dense stellar cores in their formation phase \citep[see also][]{Patel_2013, Stefanon_2013, Barro_2013, Barro_2014a, Barro_2014b, Williams_2015} or quenching progenitors suggestive of the transition stage to compact quiescent remnants \citep{Marsan_2015}.
Recent and upcoming efforts to secure samples of very distant quiescent galaxies and of their immediate progenitors, their observation with state-of-the art and new instruments to probe their stellar population, gas content, and structural and kinematical properties, and the
detailed comparison with state-of-the-art simulations, will soon provide
new constraints on the early formation of massive quiescent galaxies.

\section*{Acknowledgements}
We thank the anonymous referee for a constructive report that improved the presentation of this work. PL and VS acknowledge support from the Deutsches Zentrum für Luft- und Raumfahrt Verbundforschung grant 50OR1805. VS and MP acknowledge support from the ERC-StG ClustersXCosmo grant agreement 716762. AC acknowledges support from grant PRIN MIUR 2017 - 20173ML3WW\_001. S.J. acknowledges financial support from the Spanish Ministry of Science, Innovation and Universities (MICIU) under AYA2017-84061-P, co-financed by FEDER (European Regional Development Funds).
This research is based on observations made with the NASA/ESA Hubble Space Telescope obtained from the Space Telescope Science Institute, which is operated by the Association of Universities for Research in Astronomy, Inc., under NASA contract NAS 5–26555. These observations are associated with program ID 15229.
Partly based on data products from observations made with ESO Telescopes at the La Silla Paranal Observatory under ESO programme ID 179.A-2005 and on data products produced by TERAPIX and the Cambridge Astronomy Survey Unit on behalf of the UltraVISTA consortium.
This study is partly based on a K$_{s}$-selected catalog of the COSMOS/UltraVISTA field from \cite{Muzzin_2013}.  The catalog contains PSF-matched photometry in 30 photometric bands covering the wavelength range 0.15$\micron$ $\rightarrow$ 24$\micron$ and includes the available $GALEX$ \citep{Martin_2005}, CFHT/Subaru \citep{Capak_2007}, UltraVISTA \citep{McCracken_2012}, S-COSMOS \citep{Sanders_2007}, and zCOSMOS \citep{Lilly_2009} datasets.

\section*{Data Availability}
The data of the HST program underlying this article are available in the HST archive\footnote{https://archive.stsci.edu/hst/}. References for additional data are given in the text.



\bibliographystyle{mnras}
\bibliography{biblio} 

\begin{thebibliography}{}
\makeatletter
\relax
\def\mn@urlcharsother{\let\do\@makeother \do\$\do\&\do\#\do\^\do\_\do\%\do\~}
\def\mn@doi{\begingroup\mn@urlcharsother \@ifnextchar [ {\mn@doi@}
  {\mn@doi@[]}}
\def\mn@doi@[#1]#2{\def\@tempa{#1}\ifx\@tempa\@empty \href
  {http://dx.doi.org/#2} {doi:#2}\else \href {http://dx.doi.org/#2} {#1}\fi
  \endgroup}
\def\mn@eprint#1#2{\mn@eprint@#1:#2::\@nil}
\def\mn@eprint@arXiv#1{\href {http://arxiv.org/abs/#1} {{\tt arXiv:#1}}}
\def\mn@eprint@dblp#1{\href {http://dblp.uni-trier.de/rec/bibtex/#1.xml}
  {dblp:#1}}
\def\mn@eprint@#1:#2:#3:#4\@nil{\def\@tempa {#1}\def\@tempb {#2}\def\@tempc
  {#3}\ifx \@tempc \@empty \let \@tempc \@tempb \let \@tempb \@tempa \fi \ifx
  \@tempb \@empty \def\@tempb {arXiv}\fi \@ifundefined
  {mn@eprint@\@tempb}{\@tempb:\@tempc}{\expandafter \expandafter \csname
  mn@eprint@\@tempb\endcsname \expandafter{\@tempc}}}

\bibitem[\protect\citeauthoryear{{Almaini} et~al.,}{{Almaini}
  et~al.}{2017}]{Almaini_2017}
{Almaini} O.,  et~al., 2017, \mn@doi [\mnras] {10.1093/mnras/stx1957}, \href
  {https://ui.adsabs.harvard.edu/abs/2017MNRAS.472.1401A} {472, 1401}

\bibitem[\protect\citeauthoryear{{Athanassoula}, {Rodionov}, {Peschken}  \&
  {Lambert}}{{Athanassoula} et~al.}{2016}]{Athanassoula_2016}
{Athanassoula} E.,  {Rodionov} S.~A.,  {Peschken} N.,   {Lambert} J.~C.,  2016,
  \mn@doi [\apj] {10.3847/0004-637X/821/2/90}, \href
  {https://ui.adsabs.harvard.edu/abs/2016ApJ...821...90A} {821, 90}

\bibitem[\protect\citeauthoryear{{Barro} et~al.,}{{Barro}
  et~al.}{2013}]{Barro_2013}
{Barro} G.,  et~al., 2013, \mn@doi [\apj] {10.1088/0004-637X/765/2/104}, \href
  {https://ui.adsabs.harvard.edu/abs/2013ApJ...765..104B} {765, 104}

\bibitem[\protect\citeauthoryear{{Barro} et~al.,}{{Barro}
  et~al.}{2014a}]{Barro_2014a}
{Barro} G.,  et~al., 2014a, \mn@doi [\apj] {10.1088/0004-637X/791/1/52}, \href
  {https://ui.adsabs.harvard.edu/abs/2014ApJ...791...52B} {791, 52}

\bibitem[\protect\citeauthoryear{{Barro} et~al.,}{{Barro}
  et~al.}{2014b}]{Barro_2014b}
{Barro} G.,  et~al., 2014b, \mn@doi [\apj] {10.1088/0004-637X/795/2/145}, \href
  {https://ui.adsabs.harvard.edu/abs/2014ApJ...795..145B} {795, 145}

\bibitem[\protect\citeauthoryear{{Barro} et~al.,}{{Barro}
  et~al.}{2017}]{Barro_2017}
{Barro} G.,  et~al., 2017, \mn@doi [\apj] {10.3847/1538-4357/aa6b05}, \href
  {https://ui.adsabs.harvard.edu/abs/2017ApJ...840...47B} {840, 47}

\bibitem[\protect\citeauthoryear{{Bell} et~al.,}{{Bell}
  et~al.}{2006}]{Bell_2006}
{Bell} E.~F.,  et~al., 2006, \mn@doi [\apj] {10.1086/499931}, \href
  {https://ui.adsabs.harvard.edu/abs/2006ApJ...640..241B} {640, 241}

\bibitem[\protect\citeauthoryear{{Bell} et~al.,}{{Bell}
  et~al.}{2012}]{Bell_2012}
{Bell} E.~F.,  et~al., 2012, \mn@doi [\apj] {10.1088/0004-637X/753/2/167},
  \href {https://ui.adsabs.harvard.edu/abs/2012ApJ...753..167B} {753, 167}

\bibitem[\protect\citeauthoryear{{Belli}, {Newman}  \& {Ellis}}{{Belli}
  et~al.}{2014}]{Belli_2014}
{Belli} S.,  {Newman} A.~B.,   {Ellis} R.~S.,  2014, \mn@doi [\apj]
  {10.1088/0004-637X/783/2/117}, \href
  {https://ui.adsabs.harvard.edu/abs/2014ApJ...783..117B} {783, 117}

\bibitem[\protect\citeauthoryear{{Belli}, {Newman}  \& {Ellis}}{{Belli}
  et~al.}{2015}]{Belli_2015}
{Belli} S.,  {Newman} A.~B.,   {Ellis} R.~S.,  2015, \mn@doi [\apj]
  {10.1088/0004-637X/799/2/206}, \href
  {https://ui.adsabs.harvard.edu/abs/2015ApJ...799..206B} {799, 206}

\bibitem[\protect\citeauthoryear{{Belli}, {Newman}  \& {Ellis}}{{Belli}
  et~al.}{2017}]{Belli_2017}
{Belli} S.,  {Newman} A.~B.,   {Ellis} R.~S.,  2017, \mn@doi [\apj]
  {10.3847/1538-4357/834/1/18}, \href
  {https://ui.adsabs.harvard.edu/abs/2017ApJ...834...18B} {834, 18}

\bibitem[\protect\citeauthoryear{Belli, Newman  \& Ellis}{Belli
  et~al.}{2019}]{Belli_2019}
Belli S.,  Newman A.~B.,   Ellis R.~S.,  2019, \mn@doi [The Astrophysical
  Journal] {10.3847/1538-4357/ab07af}, 874, 17

\bibitem[\protect\citeauthoryear{{Bertin} \& {Arnouts}}{{Bertin} \&
  {Arnouts}}{1996}]{sextractor}
{Bertin} E.,  {Arnouts} S.,  1996, \mn@doi [\aaps] {10.1051/aas:1996164}, \href
  {https://ui.adsabs.harvard.edu/abs/1996A%26AS..117..393B} {117, 393}

\bibitem[\protect\citeauthoryear{{Bertin}, {Mellier}, {Radovich}, {Missonnier},
  {Didelon}  \& {Morin}}{{Bertin} et~al.}{2002}]{swarp}
{Bertin} E.,  {Mellier} Y.,  {Radovich} M.,  {Missonnier} G.,  {Didelon} P.,
  {Morin} B.,  2002, in {Bohlender} D.~A.,  {Durand} D.,   {Handley} T.~H.,
  eds,  Astronomical Society of the Pacific Conference Series Vol. 281,
  Astronomical Data Analysis Software and Systems XI. p.~228

\bibitem[\protect\citeauthoryear{Bezanson, van Dokkum, Tal, Marchesini, Kriek,
  Franx  \& Coppi}{Bezanson et~al.}{2009}]{Bezanson_2009}
Bezanson R.,  van Dokkum P.~G.,  Tal T.,  Marchesini D.,  Kriek M.,  Franx M.,
   Coppi P.,  2009, \mn@doi [The Astrophysical Journal]
  {10.1088/0004-637x/697/2/1290}, 697, 1290

\bibitem[\protect\citeauthoryear{{Bezanson}, {van Dokkum}  \&
  {Franx}}{{Bezanson} et~al.}{2012}]{Bezanson_2012}
{Bezanson} R.,  {van Dokkum} P.,   {Franx} M.,  2012, \mn@doi [\apj]
  {10.1088/0004-637X/760/1/62}, \href
  {https://ui.adsabs.harvard.edu/abs/2012ApJ...760...62B} {760, 62}

\bibitem[\protect\citeauthoryear{{Bezanson} et~al.,}{{Bezanson}
  et~al.}{2018}]{Bezanson_2018}
{Bezanson} R.,  et~al., 2018, \mn@doi [\apj] {10.3847/1538-4357/aabc55}, \href
  {https://ui.adsabs.harvard.edu/abs/2018ApJ...858...60B} {858, 60}

\bibitem[\protect\citeauthoryear{{Birnboim} \& {Dekel}}{{Birnboim} \&
  {Dekel}}{2003}]{Birnboim_2003}
{Birnboim} Y.,  {Dekel} A.,  2003, \mn@doi [\mnras]
  {10.1046/j.1365-8711.2003.06955.x}, \href
  {https://ui.adsabs.harvard.edu/abs/2003MNRAS.345..349B} {345, 349}

\bibitem[\protect\citeauthoryear{{Brammer}}{{Brammer}}{2018}]{Brammer_2018}
{Brammer} G.,  2018, {Gbrammer/Grizli: Preliminary Release},
  \mn@doi{10.5281/zenodo.1146904}

\bibitem[\protect\citeauthoryear{Brammer, van Dokkum  \& Coppi}{Brammer
  et~al.}{2008}]{Brammer_2008}
Brammer G.~B.,  van Dokkum P.~G.,   Coppi P.,  2008, \mn@doi [The Astrophysical
  Journal] {10.1086/591786}, 686, 1503

\bibitem[\protect\citeauthoryear{{Brammer} et~al.,}{{Brammer}
  et~al.}{2011}]{Brammer_2011}
{Brammer} G.~B.,  et~al., 2011, \mn@doi [\apj] {10.1088/0004-637X/739/1/24},
  \href {https://ui.adsabs.harvard.edu/abs/2011ApJ...739...24B} {739, 24}

\bibitem[\protect\citeauthoryear{{Brinchmann}, {Charlot}, {White}, {Tremonti},
  {Kauffmann}, {Heckman}  \& {Brinkmann}}{{Brinchmann}
  et~al.}{2004}]{Brinchmann_2004}
{Brinchmann} J.,  {Charlot} S.,  {White} S.~D.~M.,  {Tremonti} C.,  {Kauffmann}
  G.,  {Heckman} T.,   {Brinkmann} J.,  2004, \mn@doi [\mnras]
  {10.1111/j.1365-2966.2004.07881.x}, \href
  {https://ui.adsabs.harvard.edu/abs/2004MNRAS.351.1151B} {351, 1151}

\bibitem[\protect\citeauthoryear{Bruzual \& Charlot}{Bruzual \&
  Charlot}{2003}]{Bruzual2003}
Bruzual G.,  Charlot S.,  2003, \mn@doi [Mon. Not. Roy. Astron. Soc.]
  {10.1046/j.1365-8711.2003.06897.x}, 344, 1000

\bibitem[\protect\citeauthoryear{{Bundy} et~al.,}{{Bundy}
  et~al.}{2010}]{Bundy_2010}
{Bundy} K.,  et~al., 2010, \mn@doi [\apj] {10.1088/0004-637X/719/2/1969}, \href
  {https://ui.adsabs.harvard.edu/abs/2010ApJ...719.1969B} {719, 1969}

\bibitem[\protect\citeauthoryear{{Burkert} et~al.,}{{Burkert}
  et~al.}{2010}]{Burkert_2010}
{Burkert} A.,  et~al., 2010, \mn@doi [\apj] {10.1088/0004-637X/725/2/2324},
  \href {https://ui.adsabs.harvard.edu/abs/2010ApJ...725.2324B} {725, 2324}

\bibitem[\protect\citeauthoryear{Bédorf \& Portegies~Zwart}{Bédorf \&
  Portegies~Zwart}{2013}]{Bedorf_2013}
Bédorf J.,  Portegies~Zwart S.,  2013, \mn@doi [Monthly Notices of the Royal
  Astronomical Society] {10.1093/mnras/stt208}, 431, 767–780

\bibitem[\protect\citeauthoryear{Calzetti}{Calzetti}{2001}]{Calzetti2001}
Calzetti D.,  2001, \mn@doi [Publications of the Astronomical Society of the
  Pacific] {10.1086/324269}, 113, 1449

\bibitem[\protect\citeauthoryear{{Capak} et~al.,}{{Capak}
  et~al.}{2007}]{Capak_2007}
{Capak} P.,  et~al., 2007, \mn@doi [\apjs] {10.1086/519081}, \href
  {https://ui.adsabs.harvard.edu/abs/2007ApJS..172...99C} {172, 99}

\bibitem[\protect\citeauthoryear{{Cappellari}}{{Cappellari}}{2016}]{Cappellari_2016}
{Cappellari} M.,  2016, \mn@doi [\araa] {10.1146/annurev-astro-082214-122432},
  \href {https://ui.adsabs.harvard.edu/abs/2016ARA&A..54..597C} {54, 597}

\bibitem[\protect\citeauthoryear{{Carollo} et~al.,}{{Carollo}
  et~al.}{2013}]{Carollo_2013}
{Carollo} C.~M.,  et~al., 2013, \mn@doi [\apj] {10.1088/0004-637X/773/2/112},
  \href {https://ui.adsabs.harvard.edu/abs/2013ApJ...773..112C} {773, 112}

\bibitem[\protect\citeauthoryear{{Cassata} et~al.,}{{Cassata}
  et~al.}{2013}]{Cassata_2013}
{Cassata} P.,  et~al., 2013, \mn@doi [\apj] {10.1088/0004-637X/775/2/106},
  \href {https://ui.adsabs.harvard.edu/abs/2013ApJ...775..106C} {775, 106}

\bibitem[\protect\citeauthoryear{Chabrier}{Chabrier}{2003}]{Chabrier2003}
Chabrier G.,  2003, \mn@doi [Publications of the Astronomical Society of the
  Pacific] {10.1086/376392}, 115, 763

\bibitem[\protect\citeauthoryear{{Chang} et~al.,}{{Chang}
  et~al.}{2013}]{Chang_2013}
{Chang} Y.-Y.,  et~al., 2013, \mn@doi [\apj] {10.1088/0004-637X/773/2/149},
  \href {https://ui.adsabs.harvard.edu/abs/2013ApJ...773..149C} {773, 149}

\bibitem[\protect\citeauthoryear{{Cimatti} et~al.,}{{Cimatti}
  et~al.}{2004}]{Cimatti_2004}
{Cimatti} A.,  et~al., 2004, \mn@doi [\nat] {10.1038/nature02668}, \href
  {https://ui.adsabs.harvard.edu/abs/2004Natur.430..184C} {430, 184}

\bibitem[\protect\citeauthoryear{{Cimatti} et~al.,}{{Cimatti}
  et~al.}{2008}]{Cimatti_2008}
{Cimatti} A.,  et~al., 2008, \mn@doi [\aap] {10.1051/0004-6361:20078739}, \href
  {https://ui.adsabs.harvard.edu/abs/2008A&A...482...21C} {482, 21}

\bibitem[\protect\citeauthoryear{{Cimatti}, {Nipoti}  \& {Cassata}}{{Cimatti}
  et~al.}{2012}]{Cimatti_2012}
{Cimatti} A.,  {Nipoti} C.,   {Cassata} P.,  2012, \mn@doi [\mnras]
  {10.1111/j.1745-3933.2012.01237.x}, \href
  {https://ui.adsabs.harvard.edu/abs/2012MNRAS.422L..62C} {422, L62}

\bibitem[\protect\citeauthoryear{{Ciotti} \& {Ostriker}}{{Ciotti} \&
  {Ostriker}}{2007}]{Ciotti_2007}
{Ciotti} L.,  {Ostriker} J.~P.,  2007, \mn@doi [\apj] {10.1086/519833}, \href
  {https://ui.adsabs.harvard.edu/abs/2007ApJ...665.1038C} {665, 1038}

\bibitem[\protect\citeauthoryear{{Conroy}}{{Conroy}}{2013}]{Conroy_2013}
{Conroy} C.,  2013, \mn@doi [\araa] {10.1146/annurev-astro-082812-141017},
  \href {https://ui.adsabs.harvard.edu/abs/2013ARA&A..51..393C} {51, 393}

\bibitem[\protect\citeauthoryear{{D'Eugenio} et~al.,}{{D'Eugenio}
  et~al.}{2020}]{Deugenio_2020}
{D'Eugenio} C.,  et~al., 2020, \mn@doi [\apjl] {10.3847/2041-8213/ab7a96},
  \href {https://ui.adsabs.harvard.edu/abs/2020ApJ...892L...2D} {892, L2}

\bibitem[\protect\citeauthoryear{Daddi, Cimatti, Renzini, Fontana, Mignoli,
  Pozzetti, Tozzi  \& Zamorani}{Daddi et~al.}{2004}]{Daddi2004}
Daddi E.,  Cimatti A.,  Renzini A.,  Fontana A.,  Mignoli M.,  Pozzetti L.,
  Tozzi P.,   Zamorani G.,  2004, \mn@doi [The Astrophysical Journal]
  {10.1086/425569}, 617, 746

\bibitem[\protect\citeauthoryear{Daddi et~al.,}{Daddi
  et~al.}{2005}]{Daddi_2005}
Daddi E.,  et~al., 2005, \mn@doi [The Astrophysical Journal] {10.1086/430104},
  626, 680

\bibitem[\protect\citeauthoryear{{Dekel} \& {Birnboim}}{{Dekel} \&
  {Birnboim}}{2006}]{Dekel_2006}
{Dekel} A.,  {Birnboim} Y.,  2006, \mn@doi [\mnras]
  {10.1111/j.1365-2966.2006.10145.x}, \href
  {https://ui.adsabs.harvard.edu/abs/2006MNRAS.368....2D} {368, 2}

\bibitem[\protect\citeauthoryear{{Dekel} \& {Birnboim}}{{Dekel} \&
  {Birnboim}}{2008}]{Dekel_2008}
{Dekel} A.,  {Birnboim} Y.,  2008, \mn@doi [\mnras]
  {10.1111/j.1365-2966.2007.12569.x}, \href
  {https://ui.adsabs.harvard.edu/abs/2008MNRAS.383..119D} {383, 119}

\bibitem[\protect\citeauthoryear{{Dekel} \& {Burkert}}{{Dekel} \&
  {Burkert}}{2014}]{Dekel_2014}
{Dekel} A.,  {Burkert} A.,  2014, \mn@doi [\mnras] {10.1093/mnras/stt2331},
  \href {https://ui.adsabs.harvard.edu/abs/2014MNRAS.438.1870D} {438, 1870}

\bibitem[\protect\citeauthoryear{{Dekel} \& {Silk}}{{Dekel} \&
  {Silk}}{1986}]{Dekel_1986}
{Dekel} A.,  {Silk} J.,  1986, \mn@doi [\apj] {10.1086/164050}, \href
  {https://ui.adsabs.harvard.edu/abs/1986ApJ...303...39D} {303, 39}

\bibitem[\protect\citeauthoryear{{Dekel} et~al.,}{{Dekel}
  et~al.}{2009a}]{Dekel_2009a}
{Dekel} A.,  et~al., 2009a, \mn@doi [\nat] {10.1038/nature07648}, \href
  {https://ui.adsabs.harvard.edu/abs/2009Natur.457..451D} {457, 451}

\bibitem[\protect\citeauthoryear{{Dekel}, {Sari}  \& {Ceverino}}{{Dekel}
  et~al.}{2009b}]{Dekel_2009b}
{Dekel} A.,  {Sari} R.,   {Ceverino} D.,  2009b, \mn@doi [\apj]
  {10.1088/0004-637X/703/1/785}, \href
  {https://ui.adsabs.harvard.edu/abs/2009ApJ...703..785D} {703, 785}

\bibitem[\protect\citeauthoryear{{Dekel}, {Zolotov}, {Tweed}, {Cacciato},
  {Ceverino}  \& {Primack}}{{Dekel} et~al.}{2013}]{Dekel_2013}
{Dekel} A.,  {Zolotov} A.,  {Tweed} D.,  {Cacciato} M.,  {Ceverino} D.,
  {Primack} J.~R.,  2013, \mn@doi [\mnras] {10.1093/mnras/stt1338}, \href
  {https://ui.adsabs.harvard.edu/abs/2013MNRAS.435..999D} {435, 999}

\bibitem[\protect\citeauthoryear{{Emsellem} et~al.,}{{Emsellem}
  et~al.}{2011}]{Emsellem_2011}
{Emsellem} E.,  et~al., 2011, \mn@doi [\mnras]
  {10.1111/j.1365-2966.2011.18496.x}, \href
  {https://ui.adsabs.harvard.edu/abs/2011MNRAS.414..888E} {414, 888}

\bibitem[\protect\citeauthoryear{{Esdaile} et~al.,}{{Esdaile}
  et~al.}{2020}]{Esdaile_2020}
{Esdaile} J.,  et~al., 2020, arXiv e-prints, \href
  {https://ui.adsabs.harvard.edu/abs/2020arXiv201009738E} {p. arXiv:2010.09738}

\bibitem[\protect\citeauthoryear{{Feldmann} \& {Mayer}}{{Feldmann} \&
  {Mayer}}{2015}]{Feldmann_2015}
{Feldmann} R.,  {Mayer} L.,  2015, \mn@doi [\mnras] {10.1093/mnras/stu2207},
  \href {https://ui.adsabs.harvard.edu/abs/2015MNRAS.446.1939F} {446, 1939}

\bibitem[\protect\citeauthoryear{{Forrest} et~al.,}{{Forrest}
  et~al.}{2020a}]{Forrest_2020}
{Forrest} B.,  et~al., 2020a, \mn@doi [\apjl] {10.3847/2041-8213/ab5b9f}, \href
  {https://ui.adsabs.harvard.edu/abs/2020ApJ...890L...1F} {890, L1}

\bibitem[\protect\citeauthoryear{{Forrest} et~al.,}{{Forrest}
  et~al.}{2020b}]{Forrest_2020b}
{Forrest} B.,  et~al., 2020b, \mn@doi [\apj] {10.3847/1538-4357/abb819}, \href
  {https://ui.adsabs.harvard.edu/abs/2020ApJ...903...47F} {903, 47}

\bibitem[\protect\citeauthoryear{{Franx}, {van Dokkum}, {F{\"o}rster
  Schreiber}, {Wuyts}, {Labb{\'e}}  \& {Toft}}{{Franx}
  et~al.}{2008}]{Franx_2008}
{Franx} M.,  {van Dokkum} P.~G.,  {F{\"o}rster Schreiber} N.~M.,  {Wuyts} S.,
  {Labb{\'e}} I.,   {Toft} S.,  2008, \mn@doi [\apj] {10.1086/592431}, \href
  {https://ui.adsabs.harvard.edu/abs/2008ApJ...688..770F} {688, 770}

\bibitem[\protect\citeauthoryear{{Genzel}, {Baker}, {Tacconi}, {Lutz}, {Cox},
  {Guilloteau}  \& {Omont}}{{Genzel} et~al.}{2003}]{Genzel_2003}
{Genzel} R.,  {Baker} A.~J.,  {Tacconi} L.~J.,  {Lutz} D.,  {Cox} P.,
  {Guilloteau} S.,   {Omont} A.,  2003, \mn@doi [\apj] {10.1086/345718}, \href
  {https://ui.adsabs.harvard.edu/abs/2003ApJ...584..633G} {584, 633}

\bibitem[\protect\citeauthoryear{{Gilli} et~al.,}{{Gilli}
  et~al.}{2014}]{Gilli_2014}
{Gilli} R.,  et~al., 2014, \mn@doi [\aap] {10.1051/0004-6361/201322892}, \href
  {https://ui.adsabs.harvard.edu/abs/2014A&A...562A..67G} {562, A67}

\bibitem[\protect\citeauthoryear{Glazebrook et~al.,}{Glazebrook
  et~al.}{2004}]{Glazebrook_2004}
Glazebrook K.,  et~al., 2004, \mn@doi [Nature] {10.1038/nature02667}, 430, 181

\bibitem[\protect\citeauthoryear{Glazebrook et~al.,}{Glazebrook
  et~al.}{2017}]{Glazebrook_2017}
Glazebrook K.,  et~al., 2017, \mn@doi [Nature] {10.1038/nature21680}, 544, 71

\bibitem[\protect\citeauthoryear{Gobat et~al.,}{Gobat
  et~al.}{2012}]{Gobat_2012}
Gobat R.,  et~al., 2012, \mn@doi [The Astrophysical Journal]
  {10.1088/2041-8205/759/2/l44}, 759, L44

\bibitem[\protect\citeauthoryear{{Gobat} et~al.,}{{Gobat}
  et~al.}{2017}]{Gobat_2017}
{Gobat} R.,  et~al., 2017, \mn@doi [\aap] {10.1051/0004-6361/201629852}, \href
  {https://ui.adsabs.harvard.edu/abs/2017A&A...599A..95G} {599, A95}

\bibitem[\protect\citeauthoryear{{G{\'o}mez-Guijarro}
  et~al.,}{{G{\'o}mez-Guijarro} et~al.}{2019}]{Gomez-Guijarro_2019}
{G{\'o}mez-Guijarro} C.,  et~al., 2019, \mn@doi [\apj]
  {10.3847/1538-4357/ab418b}, \href
  {https://ui.adsabs.harvard.edu/abs/2019ApJ...886...88G} {886, 88}

\bibitem[\protect\citeauthoryear{Grogin et~al.,}{Grogin
  et~al.}{2011}]{Grogin2011_CandelsDesign}
Grogin N.~A.,  et~al., 2011, \mn@doi [The Astrophysical Journal Supplement
  Series] {10.1088/0067-0049/197/2/35}, 197, 35

\bibitem[\protect\citeauthoryear{Guo et~al.,}{Guo et~al.}{2009}]{Guo_2009}
Guo Y.,  et~al., 2009, \mn@doi [Monthly Notices of the Royal Astronomical
  Society] {10.1111/j.1365-2966.2009.15223.x}, 398, 1129

\bibitem[\protect\citeauthoryear{Hill, Muzzin, Franx  \& van~de Sande}{Hill
  et~al.}{2016}]{Hill_2016}
Hill A.~R.,  Muzzin A.,  Franx M.,   van~de Sande J.,  2016, \mn@doi [The
  Astrophysical Journal] {10.3847/0004-637x/819/1/74}, 819, 74

\bibitem[\protect\citeauthoryear{Hill, van~der Wel, Franx, Muzzin, Skelton,
  Momcheva, van Dokkum  \& Whitaker}{Hill et~al.}{2019}]{Hill_2019}
Hill A.~R.,  van~der Wel A.,  Franx M.,  Muzzin A.,  Skelton R.~E.,  Momcheva
  I.,  van Dokkum P.,   Whitaker K.~E.,  2019, \mn@doi [The Astrophysical
  Journal] {10.3847/1538-4357/aaf50a}, 871, 76

\bibitem[\protect\citeauthoryear{{Holden}, {van der Wel}, {Rix}  \&
  {Franx}}{{Holden} et~al.}{2012}]{Holden_2012}
{Holden} B.~P.,  {van der Wel} A.,  {Rix} H.-W.,   {Franx} M.,  2012, \mn@doi
  [\apj] {10.1088/0004-637X/749/2/96}, \href
  {https://ui.adsabs.harvard.edu/abs/2012ApJ...749...96H} {749, 96}

\bibitem[\protect\citeauthoryear{{Hopkins}, {Hernquist}, {Cox}, {Di Matteo},
  {Robertson}  \& {Springel}}{{Hopkins} et~al.}{2006}]{Hopkins_2006}
{Hopkins} P.~F.,  {Hernquist} L.,  {Cox} T.~J.,  {Di Matteo} T.,  {Robertson}
  B.,   {Springel} V.,  2006, \mn@doi [\apjs] {10.1086/499298}, \href
  {https://ui.adsabs.harvard.edu/abs/2006ApJS..163....1H} {163, 1}

\bibitem[\protect\citeauthoryear{{Hopkins}, {Cox}, {Kere{\v{s}}}  \&
  {Hernquist}}{{Hopkins} et~al.}{2008}]{Hopkins_2008}
{Hopkins} P.~F.,  {Cox} T.~J.,  {Kere{\v{s}}} D.,   {Hernquist} L.,  2008,
  \mn@doi [\apjs] {10.1086/524363}, \href
  {https://ui.adsabs.harvard.edu/abs/2008ApJS..175..390H} {175, 390}

\bibitem[\protect\citeauthoryear{{Hsu}, {Stockton}  \& {Shih}}{{Hsu}
  et~al.}{2014}]{Hsu_2014}
{Hsu} L.-Y.,  {Stockton} A.,   {Shih} H.-Y.,  2014, \mn@doi [\apj]
  {10.1088/0004-637X/796/2/92}, \href
  {https://ui.adsabs.harvard.edu/abs/2014ApJ...796...92H} {796, 92}

\bibitem[\protect\citeauthoryear{{Ilbert} et~al.,}{{Ilbert}
  et~al.}{2010}]{Ilbert_2010}
{Ilbert} O.,  et~al., 2010, \mn@doi [\apj] {10.1088/0004-637X/709/2/644}, \href
  {https://ui.adsabs.harvard.edu/abs/2010ApJ...709..644I} {709, 644}

\bibitem[\protect\citeauthoryear{{Ilbert} et~al.,}{{Ilbert}
  et~al.}{2013}]{Ilbert_2013}
{Ilbert} O.,  et~al., 2013, \mn@doi [\aap] {10.1051/0004-6361/201321100}, \href
  {https://ui.adsabs.harvard.edu/abs/2013A&A...556A..55I} {556, A55}

\bibitem[\protect\citeauthoryear{{Jin} et~al.,}{{Jin} et~al.}{2018}]{Jin_2018}
{Jin} S.,  et~al., 2018, \mn@doi [\apj] {10.3847/1538-4357/aad4af}, \href
  {https://ui.adsabs.harvard.edu/abs/2018ApJ...864...56J} {864, 56}

\bibitem[\protect\citeauthoryear{{Jin} et~al.,}{{Jin} et~al.}{2019}]{Jin_2019}
{Jin} S.,  et~al., 2019, \mn@doi [\apj] {10.3847/1538-4357/ab55d6}, \href
  {https://ui.adsabs.harvard.edu/abs/2019ApJ...887..144J} {887, 144}

\bibitem[\protect\citeauthoryear{{Kauffmann} et~al.,}{{Kauffmann}
  et~al.}{2003}]{Kauffmann_2003}
{Kauffmann} G.,  et~al., 2003, \mn@doi [\mnras]
  {10.1046/j.1365-8711.2003.06292.x}, \href
  {https://ui.adsabs.harvard.edu/abs/2003MNRAS.341...54K} {341, 54}

\bibitem[\protect\citeauthoryear{{Kere{\v{s}}}, {Katz}, {Weinberg}  \&
  {Dav{\'e}}}{{Kere{\v{s}}} et~al.}{2005}]{Keres_2005}
{Kere{\v{s}}} D.,  {Katz} N.,  {Weinberg} D.~H.,   {Dav{\'e}} R.,  2005,
  \mn@doi [\mnras] {10.1111/j.1365-2966.2005.09451.x}, \href
  {https://ui.adsabs.harvard.edu/abs/2005MNRAS.363....2K} {363, 2}

\bibitem[\protect\citeauthoryear{{Khochfar} \& {Ostriker}}{{Khochfar} \&
  {Ostriker}}{2008}]{Khochfar_2008}
{Khochfar} S.,  {Ostriker} J.~P.,  2008, \mn@doi [\apj] {10.1086/587470}, \href
  {https://ui.adsabs.harvard.edu/abs/2008ApJ...680...54K} {680, 54}

\bibitem[\protect\citeauthoryear{{Khochfar} \& {Silk}}{{Khochfar} \&
  {Silk}}{2006}]{Khochfar_2006}
{Khochfar} S.,  {Silk} J.,  2006, \mn@doi [\apjl] {10.1086/507768}, \href
  {https://ui.adsabs.harvard.edu/abs/2006ApJ...648L..21K} {648, L21}

\bibitem[\protect\citeauthoryear{{Koekemoer} et~al.,}{{Koekemoer}
  et~al.}{2007}]{Koekemoer_2007}
{Koekemoer} A.~M.,  et~al., 2007, \mn@doi [\apjs] {10.1086/520086}, \href
  {https://ui.adsabs.harvard.edu/abs/2007ApJS..172..196K} {172, 196}

\bibitem[\protect\citeauthoryear{Koekemoer et~al.,}{Koekemoer
  et~al.}{2011}]{Koekemoer2011_CandelsData}
Koekemoer A.~M.,  et~al., 2011, \mn@doi [The Astrophysical Journal Supplement
  Series] {10.1088/0067-0049/197/2/36}, 197, 36

\bibitem[\protect\citeauthoryear{Kriek et~al.,}{Kriek
  et~al.}{2006}]{Kriek_2006}
Kriek M.,  et~al., 2006, \mn@doi [The Astrophysical Journal] {10.1086/508371},
  649, L71

\bibitem[\protect\citeauthoryear{{Kriek}, {van Dokkum}, {Labb{\'e}}, {Franx},
  {Illingworth}, {Marchesini}  \& {Quadri}}{{Kriek}
  et~al.}{2009}]{KriekFAST2009}
{Kriek} M.,  {van Dokkum} P.~G.,  {Labb{\'e}} I.,  {Franx} M.,  {Illingworth}
  G.~D.,  {Marchesini} D.,   {Quadri} R.~F.,  2009, \mn@doi [\apj]
  {10.1088/0004-637X/700/1/221}, \href
  {https://ui.adsabs.harvard.edu/abs/2009ApJ...700..221K} {700, 221}

\bibitem[\protect\citeauthoryear{{Krogager}, {Zirm}, {Toft}, {Man}  \&
  {Brammer}}{{Krogager} et~al.}{2014}]{Krogager_2014}
{Krogager} J.~K.,  {Zirm} A.~W.,  {Toft} S.,  {Man} A.,   {Brammer} G.,  2014,
  \mn@doi [\apj] {10.1088/0004-637X/797/1/17}, \href
  {https://ui.adsabs.harvard.edu/abs/2014ApJ...797...17K} {797, 17}

\bibitem[\protect\citeauthoryear{Kubo, Yamada, Ichikawa, Kajisawa, Matsuda,
  Tanaka  \& Umehata}{Kubo et~al.}{2017}]{Kubo2017}
Kubo M.,  Yamada T.,  Ichikawa T.,  Kajisawa M.,  Matsuda Y.,  Tanaka I.,
  Umehata H.,  2017, \mn@doi [Monthly Notices of the Royal Astronomical
  Society] {10.1093/mnras/stx920}, 469, 2235

\bibitem[\protect\citeauthoryear{Kubo, Tanaka, Yabe, Toft, Stockmann  \&
  G{\'{o}}mez-Guijarro}{Kubo et~al.}{2018}]{Kubo2018}
Kubo M.,  Tanaka M.,  Yabe K.,  Toft S.,  Stockmann M.,   G{\'{o}}mez-Guijarro
  C.,  2018, \mn@doi [The Astrophysical Journal] {10.3847/1538-4357/aae3e8},
  867, 1

\bibitem[\protect\citeauthoryear{{Labb{\'e}} et~al.,}{{Labb{\'e}}
  et~al.}{2005}]{Labbe_2005}
{Labb{\'e}} I.,  et~al., 2005, \mn@doi [\apjl] {10.1086/430700}, \href
  {https://ui.adsabs.harvard.edu/abs/2005ApJ...624L..81L} {624, L81}

\bibitem[\protect\citeauthoryear{{Lagos}, {da Cunha}, {Robotham}, {Obreschkow},
  {Valentino}, {Fujimoto}, {Magdis}  \& {Tobar}}{{Lagos}
  et~al.}{2020}]{Lagos_2020}
{Lagos} C. d.~P.,  {da Cunha} E.,  {Robotham} A. S.~G.,  {Obreschkow} D.,
  {Valentino} F.,  {Fujimoto} S.,  {Magdis} G.~E.,   {Tobar} R.,  2020, \mn@doi
  [\mnras] {10.1093/mnras/staa2861}, \href
  {https://ui.adsabs.harvard.edu/abs/2020MNRAS.499.1948L} {499, 1948}

\bibitem[\protect\citeauthoryear{Laigle et~al.,}{Laigle
  et~al.}{2016}]{LaigleCOSMOS2016}
Laigle C.,  et~al., 2016, \mn@doi [The Astrophysical Journal Supplement Series]
  {10.3847/0067-0049/224/2/24}, 224, 24

\bibitem[\protect\citeauthoryear{{Lang} et~al.,}{{Lang}
  et~al.}{2014}]{Lang_2014}
{Lang} P.,  et~al., 2014, \mn@doi [\apj] {10.1088/0004-637X/788/1/11}, \href
  {https://ui.adsabs.harvard.edu/abs/2014ApJ...788...11L} {788, 11}

\bibitem[\protect\citeauthoryear{{Le Floc'h} et~al.,}{{Le Floc'h}
  et~al.}{2009}]{Le_Floc_h_2009}
{Le Floc'h} E.,  et~al., 2009, \mn@doi [\apj] {10.1088/0004-637X/703/1/222},
  \href {https://ui.adsabs.harvard.edu/abs/2009ApJ...703..222L} {703, 222}

\bibitem[\protect\citeauthoryear{{Lilly} \& {Carollo}}{{Lilly} \&
  {Carollo}}{2016}]{Lilly_2016}
{Lilly} S.~J.,  {Carollo} C.~M.,  2016, \mn@doi [\apj]
  {10.3847/0004-637X/833/1/1}, \href
  {https://ui.adsabs.harvard.edu/abs/2016ApJ...833....1L} {833, 1}

\bibitem[\protect\citeauthoryear{{Lilly}, {Eales}, {Gear}, {Hammer}, {Le
  F{\`e}vre}, {Crampton}, {Bond}  \& {Dunne}}{{Lilly}
  et~al.}{1999}]{Lilly_1999}
{Lilly} S.~J.,  {Eales} S.~A.,  {Gear} W. K.~P.,  {Hammer} F.,  {Le F{\`e}vre}
  O.,  {Crampton} D.,  {Bond} J.~R.,   {Dunne} L.,  1999, \mn@doi [\apj]
  {10.1086/307310}, \href
  {https://ui.adsabs.harvard.edu/abs/1999ApJ...518..641L} {518, 641}

\bibitem[\protect\citeauthoryear{{Lilly} et~al.,}{{Lilly}
  et~al.}{2009}]{Lilly_2009}
{Lilly} S.~J.,  et~al., 2009, \mn@doi [\apjs] {10.1088/0067-0049/184/2/218},
  \href {https://ui.adsabs.harvard.edu/abs/2009ApJS..184..218L} {184, 218}

\bibitem[\protect\citeauthoryear{{Lin} et~al.,}{{Lin} et~al.}{2007}]{Lin_2007}
{Lin} L.,  et~al., 2007, \mn@doi [\apjl] {10.1086/517919}, \href
  {https://ui.adsabs.harvard.edu/abs/2007ApJ...660L..51L} {660, L51}

\bibitem[\protect\citeauthoryear{{Lin} et~al.,}{{Lin} et~al.}{2008}]{Lin_2008}
{Lin} L.,  et~al., 2008, \mn@doi [\apj] {10.1086/587928}, \href
  {https://ui.adsabs.harvard.edu/abs/2008ApJ...681..232L} {681, 232}

\bibitem[\protect\citeauthoryear{{Longhetti} \& {Saracco}}{{Longhetti} \&
  {Saracco}}{2009}]{Longhetti_2009}
{Longhetti} M.,  {Saracco} P.,  2009, \mn@doi [\mnras]
  {10.1111/j.1365-2966.2008.14375.x}, \href
  {https://ui.adsabs.harvard.edu/abs/2009MNRAS.394..774L} {394, 774}

\bibitem[\protect\citeauthoryear{{Maltby}, {Almaini}, {Wild}, {Hatch},
  {Hartley}, {Simpson}, {Rowlands}  \& {Socolovsky}}{{Maltby}
  et~al.}{2018}]{Maltby_2018}
{Maltby} D.~T.,  {Almaini} O.,  {Wild} V.,  {Hatch} N.~A.,  {Hartley} W.~G.,
  {Simpson} C.,  {Rowlands} K.,   {Socolovsky} M.,  2018, \mn@doi [\mnras]
  {10.1093/mnras/sty1794}, \href
  {https://ui.adsabs.harvard.edu/abs/2018MNRAS.480..381M} {480, 381}

\bibitem[\protect\citeauthoryear{{Man} \& {Belli}}{{Man} \&
  {Belli}}{2018}]{Man_2018}
{Man} A.,  {Belli} S.,  2018, \mn@doi [Nature Astronomy]
  {10.1038/s41550-018-0558-1}, \href
  {https://ui.adsabs.harvard.edu/abs/2018NatAs...2..695M} {2, 695}

\bibitem[\protect\citeauthoryear{{Maraston}, {Daddi}, {Renzini}, {Cimatti},
  {Dickinson}, {Papovich}, {Pasquali}  \& {Pirzkal}}{{Maraston}
  et~al.}{2006}]{Maraston_2006}
{Maraston} C.,  {Daddi} E.,  {Renzini} A.,  {Cimatti} A.,  {Dickinson} M.,
  {Papovich} C.,  {Pasquali} A.,   {Pirzkal} N.,  2006, \mn@doi [\apj]
  {10.1086/508143}, \href
  {https://ui.adsabs.harvard.edu/abs/2006ApJ...652...85M} {652, 85}

\bibitem[\protect\citeauthoryear{Marchesi et~al.,}{Marchesi
  et~al.}{2016}]{Marchesi_2016}
Marchesi S.,  et~al., 2016, \mn@doi [The Astrophysical Journal]
  {10.3847/0004-637x/817/1/34}, 817, 34

\bibitem[\protect\citeauthoryear{{Marchesini} et~al.,}{{Marchesini}
  et~al.}{2010}]{Marchesini_2010}
{Marchesini} D.,  et~al., 2010, \mn@doi [\apj] {10.1088/0004-637X/725/1/1277},
  \href {https://ui.adsabs.harvard.edu/abs/2010ApJ...725.1277M} {725, 1277}

\bibitem[\protect\citeauthoryear{{Marchesini} et~al.,}{{Marchesini}
  et~al.}{2014}]{Marchesini_2014}
{Marchesini} D.,  et~al., 2014, \mn@doi [\apj] {10.1088/0004-637X/794/1/65},
  \href {https://ui.adsabs.harvard.edu/abs/2014ApJ...794...65M} {794, 65}

\bibitem[\protect\citeauthoryear{{Marleau} \& {Simard}}{{Marleau} \&
  {Simard}}{1998}]{Marleau_1998}
{Marleau} F.~R.,  {Simard} L.,  1998, \mn@doi [\apj] {10.1086/306356}, \href
  {https://ui.adsabs.harvard.edu/abs/1998ApJ...507..585M} {507, 585}

\bibitem[\protect\citeauthoryear{{Marsan} et~al.,}{{Marsan}
  et~al.}{2015}]{Marsan_2015}
{Marsan} Z.~C.,  et~al., 2015, \mn@doi [\apj] {10.1088/0004-637X/801/2/133},
  \href {https://ui.adsabs.harvard.edu/abs/2015ApJ...801..133M} {801, 133}

\bibitem[\protect\citeauthoryear{{Marsan}, {Marchesini}, {Brammer}, {Geier},
  {Kado-Fong}, {Labb{\'e}}, {Muzzin}  \& {Stefanon}}{{Marsan}
  et~al.}{2017}]{Marsan_2017}
{Marsan} Z.~C.,  {Marchesini} D.,  {Brammer} G.~B.,  {Geier} S.,  {Kado-Fong}
  E.,  {Labb{\'e}} I.,  {Muzzin} A.,   {Stefanon} M.,  2017, \mn@doi [\apj]
  {10.3847/1538-4357/aa7206}, \href
  {https://ui.adsabs.harvard.edu/abs/2017ApJ...842...21M} {842, 21}

\bibitem[\protect\citeauthoryear{{Marsan} et~al.,}{{Marsan}
  et~al.}{2019}]{Marsan_2019}
{Marsan} Z.~C.,  et~al., 2019, \mn@doi [\apj] {10.3847/1538-4357/aaf808}, \href
  {https://ui.adsabs.harvard.edu/abs/2019ApJ...871..201M} {871, 201}

\bibitem[\protect\citeauthoryear{{Martig}, {Bournaud}, {Teyssier}  \&
  {Dekel}}{{Martig} et~al.}{2009}]{Martig_2009}
{Martig} M.,  {Bournaud} F.,  {Teyssier} R.,   {Dekel} A.,  2009, \mn@doi
  [\apj] {10.1088/0004-637X/707/1/250}, \href
  {https://ui.adsabs.harvard.edu/abs/2009ApJ...707..250M} {707, 250}

\bibitem[\protect\citeauthoryear{{Martin} et~al.,}{{Martin}
  et~al.}{2005}]{Martin_2005}
{Martin} D.~C.,  et~al., 2005, \mn@doi [\apjl] {10.1086/426387}, \href
  {https://ui.adsabs.harvard.edu/abs/2005ApJ...619L...1M} {619, L1}

\bibitem[\protect\citeauthoryear{{Massey}, {Stoughton}, {Leauthaud}, {Rhodes},
  {Koekemoer}, {Ellis}  \& {Shaghoulian}}{{Massey} et~al.}{2010}]{Massey_2010}
{Massey} R.,  {Stoughton} C.,  {Leauthaud} A.,  {Rhodes} J.,  {Koekemoer} A.,
  {Ellis} R.,   {Shaghoulian} E.,  2010, \mn@doi [\mnras]
  {10.1111/j.1365-2966.2009.15638.x}, \href
  {https://ui.adsabs.harvard.edu/abs/2010MNRAS.401..371M} {401, 371}

\bibitem[\protect\citeauthoryear{{McCracken} et~al.,}{{McCracken}
  et~al.}{2010}]{McCracken_2010}
{McCracken} H.~J.,  et~al., 2010, \mn@doi [\apj] {10.1088/0004-637X/708/1/202},
  \href {https://ui.adsabs.harvard.edu/abs/2010ApJ...708..202M} {708, 202}

\bibitem[\protect\citeauthoryear{{McCracken} et~al.,}{{McCracken}
  et~al.}{2012}]{McCracken_2012}
{McCracken} H.~J.,  et~al., 2012, \mn@doi [\aap] {10.1051/0004-6361/201219507},
  \href {https://ui.adsabs.harvard.edu/abs/2012A&A...544A.156M} {544, A156}

\bibitem[\protect\citeauthoryear{{McGrath}, {Stockton}, {Canalizo}, {Iye}  \&
  {Maihara}}{{McGrath} et~al.}{2008}]{McGrath_2008}
{McGrath} E.~J.,  {Stockton} A.,  {Canalizo} G.,  {Iye} M.,   {Maihara} T.,
  2008, \mn@doi [\apj] {10.1086/589631}, \href
  {https://ui.adsabs.harvard.edu/abs/2008ApJ...682..303M} {682, 303}

\bibitem[\protect\citeauthoryear{{McLure} et~al.,}{{McLure}
  et~al.}{2013}]{McLure_2013}
{McLure} R.~J.,  et~al., 2013, \mn@doi [\mnras] {10.1093/mnras/sts092}, \href
  {https://ui.adsabs.harvard.edu/abs/2013MNRAS.428.1088M} {428, 1088}

\bibitem[\protect\citeauthoryear{{Merlin} et~al.,}{{Merlin}
  et~al.}{2018}]{Merlin_2018}
{Merlin} E.,  et~al., 2018, \mn@doi [\mnras] {10.1093/mnras/stx2385}, \href
  {https://ui.adsabs.harvard.edu/abs/2018MNRAS.473.2098M} {473, 2098}

\bibitem[\protect\citeauthoryear{Momcheva et~al.,}{Momcheva
  et~al.}{2016}]{Momcheva2016_3Dhst}
Momcheva I.~G.,  et~al., 2016, \mn@doi [The Astrophysical Journal Supplement
  Series] {10.3847/0067-0049/225/2/27}, 225, 27

\bibitem[\protect\citeauthoryear{Mowla et~al.,}{Mowla
  et~al.}{2019}]{Mowla_2019_b}
Mowla L.~A.,  et~al., 2019, \mn@doi [The Astrophysical Journal]
  {10.3847/1538-4357/ab290a}, 880, 57

\bibitem[\protect\citeauthoryear{{Muzzin}, {van Dokkum}, {Franx}, {Marchesini},
  {Kriek}  \& {Labb{\'e}}}{{Muzzin} et~al.}{2009}]{Muzzin_2009}
{Muzzin} A.,  {van Dokkum} P.,  {Franx} M.,  {Marchesini} D.,  {Kriek} M.,
  {Labb{\'e}} I.,  2009, \mn@doi [\apjl] {10.1088/0004-637X/706/1/L188}, \href
  {https://ui.adsabs.harvard.edu/abs/2009ApJ...706L.188M} {706, L188}

\bibitem[\protect\citeauthoryear{Muzzin et~al.,}{Muzzin
  et~al.}{2013a}]{Muzzin_2013}
Muzzin A.,  et~al., 2013a, \mn@doi [The Astrophysical Journal Supplement
  Series] {10.1088/0067-0049/206/1/8}, 206, 8

\bibitem[\protect\citeauthoryear{{Muzzin} et~al.,}{{Muzzin}
  et~al.}{2013b}]{Muzzin_2013b}
{Muzzin} A.,  et~al., 2013b, \mn@doi [\apj] {10.1088/0004-637X/777/1/18}, \href
  {https://ui.adsabs.harvard.edu/abs/2013ApJ...777...18M} {777, 18}

\bibitem[\protect\citeauthoryear{{Naab}, {Khochfar}  \& {Burkert}}{{Naab}
  et~al.}{2006}]{Naab_2006}
{Naab} T.,  {Khochfar} S.,   {Burkert} A.,  2006, \mn@doi [\apjl]
  {10.1086/500205}, \href
  {https://ui.adsabs.harvard.edu/abs/2006ApJ...636L..81N} {636, L81}

\bibitem[\protect\citeauthoryear{{Naab}, Johansson  \& Ostriker}{{Naab}
  et~al.}{2009}]{Naab_2009}
{Naab} T.,  Johansson P.~H.,   Ostriker J.~P.,  2009, \mn@doi [The
  Astrophysical Journal] {10.1088/0004-637x/699/2/l178}, 699, L178

\bibitem[\protect\citeauthoryear{{Nelson} et~al.,}{{Nelson}
  et~al.}{2014}]{Nelson_2014}
{Nelson} E.,  et~al., 2014, \mn@doi [\nat] {10.1038/nature13616}, \href
  {https://ui.adsabs.harvard.edu/abs/2014Natur.513..394N} {513, 394}

\bibitem[\protect\citeauthoryear{{Newman}, {Belli}  \& {Ellis}}{{Newman}
  et~al.}{2015}]{Newman_2015}
{Newman} A.~B.,  {Belli} S.,   {Ellis} R.~S.,  2015, \mn@doi [\apjl]
  {10.1088/2041-8205/813/1/L7}, \href
  {https://ui.adsabs.harvard.edu/abs/2015ApJ...813L...7N} {813, L7}

\bibitem[\protect\citeauthoryear{{Newman}, {Belli}, {Ellis}  \&
  {Patel}}{{Newman} et~al.}{2018}]{Newman_2018}
{Newman} A.~B.,  {Belli} S.,  {Ellis} R.~S.,   {Patel} S.~G.,  2018, \mn@doi
  [\apj] {10.3847/1538-4357/aacd4f}, \href
  {https://ui.adsabs.harvard.edu/abs/2018ApJ...862..126N} {862, 126}

\bibitem[\protect\citeauthoryear{{Onodera} et~al.,}{{Onodera}
  et~al.}{2012}]{Onodera_2012}
{Onodera} M.,  et~al., 2012, \mn@doi [\apj] {10.1088/0004-637X/755/1/26}, \href
  {https://ui.adsabs.harvard.edu/abs/2012ApJ...755...26O} {755, 26}

\bibitem[\protect\citeauthoryear{{Onodera} et~al.,}{{Onodera}
  et~al.}{2015}]{Onodera_2015}
{Onodera} M.,  et~al., 2015, \mn@doi [\apj] {10.1088/0004-637X/808/2/161},
  \href {https://ui.adsabs.harvard.edu/abs/2015ApJ...808..161O} {808, 161}

\bibitem[\protect\citeauthoryear{{Oser}, {Ostriker}, {Naab}, {Johansson}  \&
  {Burkert}}{{Oser} et~al.}{2010}]{Oser_2010}
{Oser} L.,  {Ostriker} J.~P.,  {Naab} T.,  {Johansson} P.~H.,   {Burkert} A.,
  2010, \mn@doi [\apj] {10.1088/0004-637X/725/2/2312}, \href
  {https://ui.adsabs.harvard.edu/abs/2010ApJ...725.2312O} {725, 2312}

\bibitem[\protect\citeauthoryear{{Oser}, {Naab}, {Ostriker}  \&
  {Johansson}}{{Oser} et~al.}{2012}]{Oser_2012}
{Oser} L.,  {Naab} T.,  {Ostriker} J.~P.,   {Johansson} P.~H.,  2012, \mn@doi
  [\apj] {10.1088/0004-637X/744/1/63}, \href
  {https://ui.adsabs.harvard.edu/abs/2012ApJ...744...63O} {744, 63}

\bibitem[\protect\citeauthoryear{{Oteo} et~al.,}{{Oteo}
  et~al.}{2017}]{Oteo_2017}
{Oteo} I.,  et~al., 2017, arXiv e-prints, \href
  {https://ui.adsabs.harvard.edu/abs/2017arXiv170904191O} {p. arXiv:1709.04191}

\bibitem[\protect\citeauthoryear{{Pacifici} et~al.,}{{Pacifici}
  et~al.}{2015}]{Pacifici_2015}
{Pacifici} C.,  et~al., 2015, \mn@doi [\mnras] {10.1093/mnras/stu2447}, \href
  {https://ui.adsabs.harvard.edu/abs/2015MNRAS.447..786P} {447, 786}

\bibitem[\protect\citeauthoryear{{Pannella} et~al.,}{{Pannella}
  et~al.}{2009}]{Pannella_2009}
{Pannella} M.,  et~al., 2009, \mn@doi [\apj] {10.1088/0004-637X/701/1/787},
  \href {https://ui.adsabs.harvard.edu/abs/2009ApJ...701..787P} {701, 787}

\bibitem[\protect\citeauthoryear{{Patel} et~al.,}{{Patel}
  et~al.}{2013}]{Patel_2013}
{Patel} S.~G.,  et~al., 2013, \mn@doi [\apj] {10.1088/0004-637X/766/1/15},
  \href {https://ui.adsabs.harvard.edu/abs/2013ApJ...766...15P} {766, 15}

\bibitem[\protect\citeauthoryear{{Patel}, {Hong}, {Quadri}, {Holden}  \&
  {Williams}}{{Patel} et~al.}{2017}]{Patel_2017}
{Patel} S.~G.,  {Hong} Y.~X.,  {Quadri} R.~F.,  {Holden} B.~P.,   {Williams}
  R.~J.,  2017, \mn@doi [\apj] {10.3847/1538-4357/aa6bf4}, \href
  {https://ui.adsabs.harvard.edu/abs/2017ApJ...839..127P} {839, 127}

\bibitem[\protect\citeauthoryear{{Peng} \& {Renzini}}{{Peng} \&
  {Renzini}}{2020}]{Peng_2020}
{Peng} Y.-j.,  {Renzini} A.,  2020, \mn@doi [\mnras] {10.1093/mnrasl/slz163},
  \href {https://ui.adsabs.harvard.edu/abs/2020MNRAS.491L..51P} {491, L51}

\bibitem[\protect\citeauthoryear{{Peng}, {Ho}, {Impey}  \& {Rix}}{{Peng}
  et~al.}{2002}]{galfit2002}
{Peng} C.~Y.,  {Ho} L.~C.,  {Impey} C.~D.,   {Rix} H.-W.,  2002, \mn@doi [\aj]
  {10.1086/340952}, \href
  {https://ui.adsabs.harvard.edu/abs/2002AJ....124..266P} {124, 266}

\bibitem[\protect\citeauthoryear{{Peng}, {Ho}, {Impey}  \& {Rix}}{{Peng}
  et~al.}{2010a}]{galfit2010}
{Peng} C.~Y.,  {Ho} L.~C.,  {Impey} C.~D.,   {Rix} H.-W.,  2010a, \mn@doi [\aj]
  {10.1088/0004-6256/139/6/2097}, \href
  {https://ui.adsabs.harvard.edu/abs/2010AJ....139.2097P} {139, 2097}

\bibitem[\protect\citeauthoryear{{Peng} et~al.,}{{Peng}
  et~al.}{2010b}]{Peng_2010}
{Peng} Y.-j.,  et~al., 2010b, \mn@doi [\apj] {10.1088/0004-637X/721/1/193},
  \href {https://ui.adsabs.harvard.edu/abs/2010ApJ...721..193P} {721, 193}

\bibitem[\protect\citeauthoryear{{Perez}, {Michel-Dansac}  \&
  {Tissera}}{{Perez} et~al.}{2011}]{Perez_2011}
{Perez} J.,  {Michel-Dansac} L.,   {Tissera} P.~B.,  2011, \mn@doi [\mnras]
  {10.1111/j.1365-2966.2011.19300.x}, \href
  {https://ui.adsabs.harvard.edu/abs/2011MNRAS.417..580P} {417, 580}

\bibitem[\protect\citeauthoryear{{Pignatelli}, {Fasano}  \&
  {Cassata}}{{Pignatelli} et~al.}{2006}]{Pignatelli_2006}
{Pignatelli} E.,  {Fasano} G.,   {Cassata} P.,  2006, \mn@doi [\aap]
  {10.1051/0004-6361:20041704}, \href
  {https://ui.adsabs.harvard.edu/abs/2006A&A...446..373P} {446, 373}

\bibitem[\protect\citeauthoryear{{Poggianti} et~al.,}{{Poggianti}
  et~al.}{2013}]{Poggianti_2013}
{Poggianti} B.~M.,  et~al., 2013, \mn@doi [\apj] {10.1088/0004-637X/762/2/77},
  \href {https://ui.adsabs.harvard.edu/abs/2013ApJ...762...77P} {762, 77}

\bibitem[\protect\citeauthoryear{{Renzini} et~al.,}{{Renzini}
  et~al.}{2018}]{Renzini_2018}
{Renzini} A.,  et~al., 2018, \mn@doi [\apj] {10.3847/1538-4357/aad09b}, \href
  {https://ui.adsabs.harvard.edu/abs/2018ApJ...863...16R} {863, 16}

\bibitem[\protect\citeauthoryear{{Sanders} et~al.,}{{Sanders}
  et~al.}{2007}]{Sanders_2007}
{Sanders} D.~B.,  et~al., 2007, \mn@doi [\apjs] {10.1086/517885}, \href
  {https://ui.adsabs.harvard.edu/abs/2007ApJS..172...86S} {172, 86}

\bibitem[\protect\citeauthoryear{{Saracco}, {Longhetti}  \&
  {Andreon}}{{Saracco} et~al.}{2009}]{Saracco_2009}
{Saracco} P.,  {Longhetti} M.,   {Andreon} S.,  2009, \mn@doi [\mnras]
  {10.1111/j.1365-2966.2008.14085.x}, \href
  {https://ui.adsabs.harvard.edu/abs/2009MNRAS.392..718S} {392, 718}

\bibitem[\protect\citeauthoryear{{Sargent} et~al.,}{{Sargent}
  et~al.}{2007}]{Sargent_2007}
{Sargent} M.~T.,  et~al., 2007, \mn@doi [\apjs] {10.1086/516584}, \href
  {https://ui.adsabs.harvard.edu/abs/2007ApJS..172..434S} {172, 434}

\bibitem[\protect\citeauthoryear{{Schreiber} et~al.,}{{Schreiber}
  et~al.}{2015}]{Schreiber_2015}
{Schreiber} C.,  et~al., 2015, \mn@doi [\aap] {10.1051/0004-6361/201425017},
  \href {https://ui.adsabs.harvard.edu/abs/2015A&A...575A..74S} {575, A74}

\bibitem[\protect\citeauthoryear{{Schreiber} et~al.,}{{Schreiber}
  et~al.}{2018}]{Schreiber_2018}
{Schreiber} C.,  et~al., 2018, \mn@doi [\aap] {10.1051/0004-6361/201833070},
  \href {https://ui.adsabs.harvard.edu/abs/2018A&A...618A..85S} {618, A85}

\bibitem[\protect\citeauthoryear{{S{\'e}rsic}}{{S{\'e}rsic}}{1963}]{Sersic_1963}
{S{\'e}rsic} J.~L.,  1963, Boletin de la Asociacion Argentina de Astronomia La
  Plata Argentina, \href
  {https://ui.adsabs.harvard.edu/abs/1963BAAA....6...41S} {6, 41}

\bibitem[\protect\citeauthoryear{{S{\'e}rsic}}{{S{\'e}rsic}}{1968}]{Sersic_1968}
{S{\'e}rsic} J.~L.,  1968, {Atlas de Galaxias Australes}

\bibitem[\protect\citeauthoryear{{Shen}, {Mo}, {White}, {Blanton}, {Kauffmann},
  {Voges}, {Brinkmann}  \& {Csabai}}{{Shen} et~al.}{2003}]{Shen_2003}
{Shen} S.,  {Mo} H.~J.,  {White} S. D.~M.,  {Blanton} M.~R.,  {Kauffmann} G.,
  {Voges} W.,  {Brinkmann} J.,   {Csabai} I.,  2003, \mn@doi [\mnras]
  {10.1046/j.1365-8711.2003.06740.x}, \href
  {https://ui.adsabs.harvard.edu/abs/2003MNRAS.343..978S} {343, 978}

\bibitem[\protect\citeauthoryear{{Simpson} et~al.,}{{Simpson}
  et~al.}{2014}]{Simpson_2014}
{Simpson} J.~M.,  et~al., 2014, \mn@doi [\apj] {10.1088/0004-637X/788/2/125},
  \href {https://ui.adsabs.harvard.edu/abs/2014ApJ...788..125S} {788, 125}

\bibitem[\protect\citeauthoryear{{Simpson} et~al.,}{{Simpson}
  et~al.}{2017}]{Simpson_2017}
{Simpson} J.~M.,  et~al., 2017, \mn@doi [\apj] {10.3847/1538-4357/aa65d0},
  \href {https://ui.adsabs.harvard.edu/abs/2017ApJ...839...58S} {839, 58}

\bibitem[\protect\citeauthoryear{Skelton et~al.,}{Skelton
  et~al.}{2014}]{Skelton2014}
Skelton R.~E.,  et~al., 2014, \mn@doi [The Astrophysical Journal Supplement
  Series] {10.1088/0067-0049/214/2/24}, 214, 24

\bibitem[\protect\citeauthoryear{{Stefanon}, {Marchesini}, {Rudnick}, {Brammer}
   \& {Whitaker}}{{Stefanon} et~al.}{2013}]{Stefanon_2013}
{Stefanon} M.,  {Marchesini} D.,  {Rudnick} G.~H.,  {Brammer} G.~B.,
  {Whitaker} K.~E.,  2013, \mn@doi [\apj] {10.1088/0004-637X/768/1/92}, \href
  {https://ui.adsabs.harvard.edu/abs/2013ApJ...768...92S} {768, 92}

\bibitem[\protect\citeauthoryear{{Stockmann} et~al.,}{{Stockmann}
  et~al.}{2020}]{Stockmann_2020}
{Stockmann} M.,  et~al., 2020, \mn@doi [\apj] {10.3847/1538-4357/ab5af4}, \href
  {https://ui.adsabs.harvard.edu/abs/2020ApJ...888....4S} {888, 4}

\bibitem[\protect\citeauthoryear{{Stockton}, {Canalizo}  \&
  {Maihara}}{{Stockton} et~al.}{2004}]{Stockton_2004}
{Stockton} A.,  {Canalizo} G.,   {Maihara} T.,  2004, \mn@doi [\apj]
  {10.1086/382235}, \href
  {https://ui.adsabs.harvard.edu/abs/2004ApJ...605...37S} {605, 37}

\bibitem[\protect\citeauthoryear{{Stockton}, {McGrath}, {Canalizo}, {Iye}  \&
  {Maihara}}{{Stockton} et~al.}{2008}]{Stockton_2008}
{Stockton} A.,  {McGrath} E.,  {Canalizo} G.,  {Iye} M.,   {Maihara} T.,  2008,
  \mn@doi [\apj] {10.1086/523789}, \href
  {https://ui.adsabs.harvard.edu/abs/2008ApJ...672..146S} {672, 146}

\bibitem[\protect\citeauthoryear{Straatman et~al.,}{Straatman
  et~al.}{2015}]{Straatman2015}
Straatman C. M.~S.,  et~al., 2015, \mn@doi [The Astrophysical Journal]
  {10.1088/2041-8205/808/1/l29}, 808, L29

\bibitem[\protect\citeauthoryear{{Strazzullo} et~al.,}{{Strazzullo}
  et~al.}{2015}]{Strazzullo_2015}
{Strazzullo} V.,  et~al., 2015, \mn@doi [\aap] {10.1051/0004-6361/201425038},
  \href {https://ui.adsabs.harvard.edu/abs/2015A&A...576L...6S} {576, L6}

\bibitem[\protect\citeauthoryear{Suess, Kriek, Price  \& Barro}{Suess
  et~al.}{2019a}]{Suess_2019}
Suess K.~A.,  Kriek M.,  Price S.~H.,   Barro G.,  2019a, \mn@doi [The
  Astrophysical Journal] {10.3847/1538-4357/ab1bda}, 877, 103

\bibitem[\protect\citeauthoryear{Suess, Kriek, Price  \& Barro}{Suess
  et~al.}{2019b}]{Suess_2019b}
Suess K.~A.,  Kriek M.,  Price S.~H.,   Barro G.,  2019b, \mn@doi [The
  Astrophysical Journal] {10.3847/2041-8213/ab4db3}, 885, L22

\bibitem[\protect\citeauthoryear{{Szomoru}, {Franx}, {Bouwens}, {van Dokkum},
  {Labb{\'e}}, {Illingworth}  \& {Trenti}}{{Szomoru}
  et~al.}{2011}]{Szomoru_2011}
{Szomoru} D.,  {Franx} M.,  {Bouwens} R.~J.,  {van Dokkum} P.~G.,  {Labb{\'e}}
  I.,  {Illingworth} G.~D.,   {Trenti} M.,  2011, \mn@doi [\apjl]
  {10.1088/2041-8205/735/1/L22}, \href
  {https://ui.adsabs.harvard.edu/abs/2011ApJ...735L..22S} {735, L22}

\bibitem[\protect\citeauthoryear{{Tacchella} et~al.,}{{Tacchella}
  et~al.}{2015}]{Tacchella_2015}
{Tacchella} S.,  et~al., 2015, \mn@doi [Science] {10.1126/science.1261094},
  \href {https://ui.adsabs.harvard.edu/abs/2015Sci...348..314T} {348, 314}

\bibitem[\protect\citeauthoryear{{Tacchella}, {Dekel}, {Carollo}, {Ceverino},
  {DeGraf}, {Lapiner}, {Mand elker}  \& {Primack}}{{Tacchella}
  et~al.}{2016}]{Tachella_2016}
{Tacchella} S.,  {Dekel} A.,  {Carollo} C.~M.,  {Ceverino} D.,  {DeGraf} C.,
  {Lapiner} S.,  {Mand elker} N.,   {Primack} J.~R.,  2016, \mn@doi [\mnras]
  {10.1093/mnras/stw303}, \href
  {https://ui.adsabs.harvard.edu/abs/2016MNRAS.458..242T} {458, 242}

\bibitem[\protect\citeauthoryear{{Tacconi} et~al.,}{{Tacconi}
  et~al.}{2008}]{Tacconi_2008}
{Tacconi} L.~J.,  et~al., 2008, \mn@doi [\apj] {10.1086/587168}, \href
  {https://ui.adsabs.harvard.edu/abs/2008ApJ...680..246T} {680, 246}

\bibitem[\protect\citeauthoryear{{Tanaka} et~al.,}{{Tanaka}
  et~al.}{2019}]{Tanaka_2019}
{Tanaka} M.,  et~al., 2019, \mn@doi [\apjl] {10.3847/2041-8213/ab4ff3}, \href
  {https://ui.adsabs.harvard.edu/abs/2019ApJ...885L..34T} {885, L34}

\bibitem[\protect\citeauthoryear{{Toft} et~al.,}{{Toft}
  et~al.}{2007}]{Toft_2007}
{Toft} S.,  et~al., 2007, \mn@doi [\apj] {10.1086/521810}, \href
  {https://ui.adsabs.harvard.edu/abs/2007ApJ...671..285T} {671, 285}

\bibitem[\protect\citeauthoryear{{Toft} et~al.,}{{Toft}
  et~al.}{2014}]{Toft_2014}
{Toft} S.,  et~al., 2014, \mn@doi [\apj] {10.1088/0004-637X/782/2/68}, \href
  {https://ui.adsabs.harvard.edu/abs/2014ApJ...782...68T} {782, 68}

\bibitem[\protect\citeauthoryear{{Toft} et~al.,}{{Toft}
  et~al.}{2017}]{Toft_2017}
{Toft} S.,  et~al., 2017, \mn@doi [\nat] {10.1038/nature22388}, \href
  {https://ui.adsabs.harvard.edu/abs/2017Natur.546..510T} {546, 510}

\bibitem[\protect\citeauthoryear{Trujillo et~al.,}{Trujillo
  et~al.}{2006}]{Trujillo_2006}
Trujillo I.,  et~al., 2006, \mn@doi [Monthly Notices of the Royal Astronomical
  Society: Letters] {10.1111/j.1745-3933.2006.00238.x}, 373, L36

\bibitem[\protect\citeauthoryear{{Trujillo}, {Ferreras}  \& {de La
  Rosa}}{{Trujillo} et~al.}{2011}]{Trujillo_2011}
{Trujillo} I.,  {Ferreras} I.,   {de La Rosa} I.~G.,  2011, \mn@doi [\mnras]
  {10.1111/j.1365-2966.2011.19017.x}, \href
  {https://ui.adsabs.harvard.edu/abs/2011MNRAS.415.3903T} {415, 3903}

\bibitem[\protect\citeauthoryear{{Valentino} et~al.,}{{Valentino}
  et~al.}{2020}]{Valentino_2020}
{Valentino} F.,  et~al., 2020, \mn@doi [\apj] {10.3847/1538-4357/ab64dc}, \href
  {https://ui.adsabs.harvard.edu/abs/2020ApJ...889...93V} {889, 93}

\bibitem[\protect\citeauthoryear{{Wellons} et~al.,}{{Wellons}
  et~al.}{2015}]{Wellons_2015}
{Wellons} S.,  et~al., 2015, \mn@doi [\mnras] {10.1093/mnras/stv303}, \href
  {https://ui.adsabs.harvard.edu/abs/2015MNRAS.449..361W} {449, 361}

\bibitem[\protect\citeauthoryear{{Whitaker} et~al.,}{{Whitaker}
  et~al.}{2010}]{Whitaker_2010}
{Whitaker} K.~E.,  et~al., 2010, \mn@doi [\apj] {10.1088/0004-637X/719/2/1715},
  \href {https://ui.adsabs.harvard.edu/abs/2010ApJ...719.1715W} {719, 1715}

\bibitem[\protect\citeauthoryear{{Whitaker}, {Kriek}, {van Dokkum}, {Bezanson},
  {Brammer}, {Franx}  \& {Labb{\'e}}}{{Whitaker} et~al.}{2012}]{Whitaker_2012}
{Whitaker} K.~E.,  {Kriek} M.,  {van Dokkum} P.~G.,  {Bezanson} R.,  {Brammer}
  G.,  {Franx} M.,   {Labb{\'e}} I.,  2012, \mn@doi [\apj]
  {10.1088/0004-637X/745/2/179}, \href
  {https://ui.adsabs.harvard.edu/abs/2012ApJ...745..179W} {745, 179}

\bibitem[\protect\citeauthoryear{{Whitaker} et~al.,}{{Whitaker}
  et~al.}{2017}]{Whitaker_2017}
{Whitaker} K.~E.,  et~al., 2017, \mn@doi [\apj] {10.3847/1538-4357/aa6258},
  \href {https://ui.adsabs.harvard.edu/abs/2017ApJ...838...19W} {838, 19}

\bibitem[\protect\citeauthoryear{Williams, Quadri, Franx, van Dokkum  \&
  Labb{\'{e}}}{Williams et~al.}{2009}]{Williams_2009}
Williams R.~J.,  Quadri R.~F.,  Franx M.,  van Dokkum P.,   Labb{\'{e}} I.,
  2009, \mn@doi [The Astrophysical Journal] {10.1088/0004-637x/691/2/1879},
  691, 1879

\bibitem[\protect\citeauthoryear{{Williams} et~al.,}{{Williams}
  et~al.}{2015}]{Williams_2015}
{Williams} C.~C.,  et~al., 2015, \mn@doi [\apj] {10.1088/0004-637X/800/1/21},
  \href {https://ui.adsabs.harvard.edu/abs/2015ApJ...800...21W} {800, 21}

\bibitem[\protect\citeauthoryear{{Williams} et~al.,}{{Williams}
  et~al.}{2017}]{Williams_2017}
{Williams} C.~C.,  et~al., 2017, \mn@doi [\apj] {10.3847/1538-4357/aa662f},
  \href {https://ui.adsabs.harvard.edu/abs/2017ApJ...838...94W} {838, 94}

\bibitem[\protect\citeauthoryear{{Wu} et~al.,}{{Wu} et~al.}{2018}]{Wu_2018}
{Wu} P.-F.,  et~al., 2018, \mn@doi [\apj] {10.3847/1538-4357/aae822}, \href
  {https://ui.adsabs.harvard.edu/abs/2018ApJ...868...37W} {868, 37}

\bibitem[\protect\citeauthoryear{{Wu} et~al.,}{{Wu} et~al.}{2020}]{Wu_2020}
{Wu} P.-F.,  et~al., 2020, \mn@doi [\apj] {10.3847/1538-4357/ab5fd9}, \href
  {https://ui.adsabs.harvard.edu/abs/2020ApJ...888...77W} {888, 77}

\bibitem[\protect\citeauthoryear{{Wuyts} et~al.,}{{Wuyts}
  et~al.}{2012}]{Wuyts_2012}
{Wuyts} S.,  et~al., 2012, \mn@doi [\apj] {10.1088/0004-637X/753/2/114}, \href
  {https://ui.adsabs.harvard.edu/abs/2012ApJ...753..114W} {753, 114}

\bibitem[\protect\citeauthoryear{{Yano}, {Kriek}, {van der Wel}  \&
  {Whitaker}}{{Yano} et~al.}{2016}]{Yano_2016}
{Yano} M.,  {Kriek} M.,  {van der Wel} A.,   {Whitaker} K.~E.,  2016, \mn@doi
  [\apjl] {10.3847/2041-8205/817/2/L21}, \href
  {https://ui.adsabs.harvard.edu/abs/2016ApJ...817L..21Y} {817, L21}

\bibitem[\protect\citeauthoryear{{Zahid} \& {Geller}}{{Zahid} \&
  {Geller}}{2017}]{Zahid_2017}
{Zahid} H.~J.,  {Geller} M.~J.,  2017, \mn@doi [\apj]
  {10.3847/1538-4357/aa7056}, \href
  {https://ui.adsabs.harvard.edu/abs/2017ApJ...841...32Z} {841, 32}

\bibitem[\protect\citeauthoryear{Zolotov et~al.,}{Zolotov
  et~al.}{2015}]{Zolotov_2015}
Zolotov A.,  et~al., 2015, \mn@doi [Monthly Notices of the Royal Astronomical
  Society] {10.1093/mnras/stv740}, 450, 2327

\bibitem[\protect\citeauthoryear{{van Dokkum} \& {Franx}}{{van Dokkum} \&
  {Franx}}{1996}]{Dokkum_1996}
{van Dokkum} P.~G.,  {Franx} M.,  1996, \mn@doi [\mnras]
  {10.1093/mnras/281.3.985}, \href
  {https://ui.adsabs.harvard.edu/abs/1996MNRAS.281..985V} {281, 985}

\bibitem[\protect\citeauthoryear{{van Dokkum} \& {Franx}}{{van Dokkum} \&
  {Franx}}{2001}]{Dokkum_2001}
{van Dokkum} P.~G.,  {Franx} M.,  2001, \mn@doi [\apj] {10.1086/320645}, \href
  {https://ui.adsabs.harvard.edu/abs/2001ApJ...553...90V} {553, 90}

\bibitem[\protect\citeauthoryear{{van Dokkum} et~al.,}{{van Dokkum}
  et~al.}{2008}]{VanDokkum_2008}
{van Dokkum} P.~G.,  et~al., 2008, \mn@doi [\apjl] {10.1086/587874}, \href
  {https://ui.adsabs.harvard.edu/abs/2008ApJ...677L...5V} {677, L5}

\bibitem[\protect\citeauthoryear{{van Dokkum} et~al.,}{{van Dokkum}
  et~al.}{2014}]{Dokkum_2014}
{van Dokkum} P.~G.,  et~al., 2014, \mn@doi [\apj] {10.1088/0004-637X/791/1/45},
  \href {https://ui.adsabs.harvard.edu/abs/2014ApJ...791...45V} {791, 45}

\bibitem[\protect\citeauthoryear{{van Dokkum} et~al.,}{{van Dokkum}
  et~al.}{2015}]{Dokkum_2015}
{van Dokkum} P.~G.,  et~al., 2015, \mn@doi [\apj] {10.1088/0004-637X/813/1/23},
  \href {https://ui.adsabs.harvard.edu/abs/2015ApJ...813...23V} {813, 23}

\bibitem[\protect\citeauthoryear{{van der Wel}, {Bell}, {van den Bosch},
  {Gallazzi}  \& {Rix}}{{van der Wel} et~al.}{2009a}]{VanDerWel_2009_a}
{van der Wel} A.,  {Bell} E.~F.,  {van den Bosch} F.~C.,  {Gallazzi} A.,
  {Rix} H.-W.,  2009a, \mn@doi [\apj] {10.1088/0004-637X/698/2/1232}, \href
  {https://ui.adsabs.harvard.edu/abs/2009ApJ...698.1232V} {698, 1232}

\bibitem[\protect\citeauthoryear{{van der Wel}, {Rix}, {Holden}, {Bell}  \&
  {Robaina}}{{van der Wel} et~al.}{2009b}]{VanDerWel_2009}
{van der Wel} A.,  {Rix} H.-W.,  {Holden} B.~P.,  {Bell} E.~F.,   {Robaina}
  A.~R.,  2009b, \mn@doi [\apjl] {10.1088/0004-637X/706/1/L120}, \href
  {https://ui.adsabs.harvard.edu/abs/2009ApJ...706L.120V} {706, L120}

\bibitem[\protect\citeauthoryear{{van der Wel} et~al.,}{{van der Wel}
  et~al.}{2011}]{VanDerWel_2011}
{van der Wel} A.,  et~al., 2011, \mn@doi [\apj] {10.1088/0004-637X/730/1/38},
  \href {https://ui.adsabs.harvard.edu/abs/2011ApJ...730...38V} {730, 38}

\bibitem[\protect\citeauthoryear{van~der Wel et~al.,}{van~der Wel
  et~al.}{2014}]{VanDerWel2014}
van~der Wel A.,  et~al., 2014, \mn@doi [The Astrophysical Journal]
  {10.1088/0004-637x/788/1/28}, 788, 28

\makeatother
\end{thebibliography}


\appendix

\section{Comparison of spectroscopic and photometric redshifts}
\label{sec:specz_photoz}
\begin{figure*}
	\includegraphics[width=\textwidth]{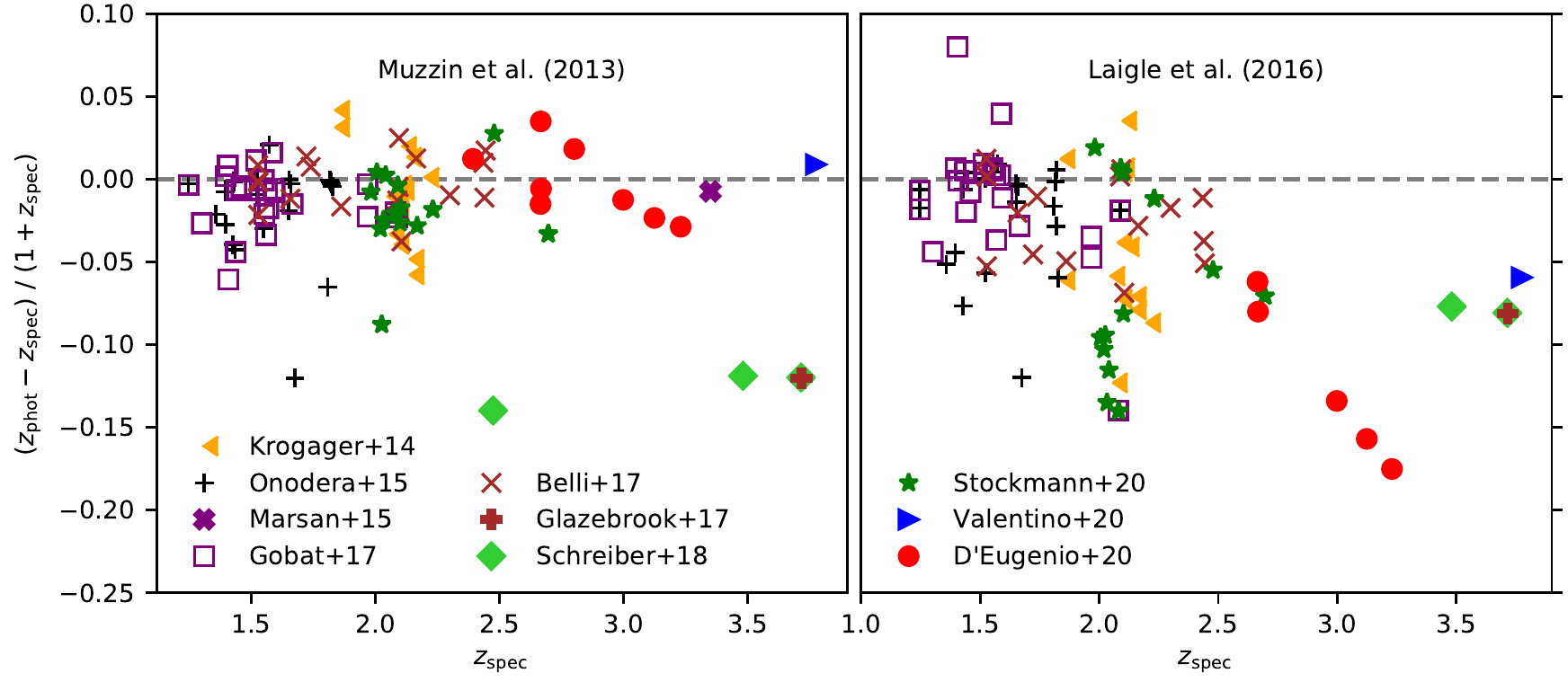}
	 \caption{Comparison of spectroscopic redshifts of quiescent galaxies with photometric redshifts from \protect\citet[][left panel]{Muzzin_2013} and \protect\citet[][right panel]{LaigleCOSMOS2016}. Sources from \protect\cite{Schreiber_2018} at $z\approx 2.5$ and \protect\citep{Marsan_2015} at $z\approx 3.4$ have a photometric redshift estimates in \protect\cite{LaigleCOSMOS2016} of $4.9$ and $z=0.3$, respectively and are not shown in the right panel.
	 The sources from \protect\cite{Glazebrook_2017} and \protect\cite{Schreiber_2018} at $z=3.7$ are the same.}
	 \label{fig:zcomparison}
\end{figure*}

In section~\ref{sec:representativeness} we compare restframe UVJ colors of our sample with the full massive parent sample at $2.5<z<3.0$. To derive UVJ colors and stellar masses for the parent sample we make use of photometric redshifts. To investigate how reliable the photometric redshifts are we compare in Figure~\ref{fig:zcomparison} spectroscopic redshifts of quiescent galaxies at $z_{\textrm{spec}}\gtrsim 1.2$ with photometric redshifts from \cite{Muzzin_2013} and \cite{LaigleCOSMOS2016}. Redshifts from \cite{Krogager_2014} and \cite{Deugenio_2020} rely on HST grism data while redshifts from \cite{Onodera_2015}, \cite{Marsan_2015}, \cite{Gobat_2017}, \cite{Belli_2017}, \cite{Glazebrook_2017}, \cite{Schreiber_2018}, \cite{Valentino_2020} and \cite{Stockmann_2020} rely on spectroscopic observations from ground-based telescopes.
Photometric redshifts from \cite{LaigleCOSMOS2016} have a larger scatter and are systematically underestimated in this redshift range, especially at $z_{\textrm{spec}}\gtrsim 2.5$. We therefore use redshifts from \cite{Muzzin_2013} together with the deeper photometry from \cite{LaigleCOSMOS2016} for our analysis in Section~\ref{sec:representativeness}.

\section{Significance of the central residuals}
\label{sec:residuals_significance}
\begin{figure*}
	\includegraphics[width=\textwidth]{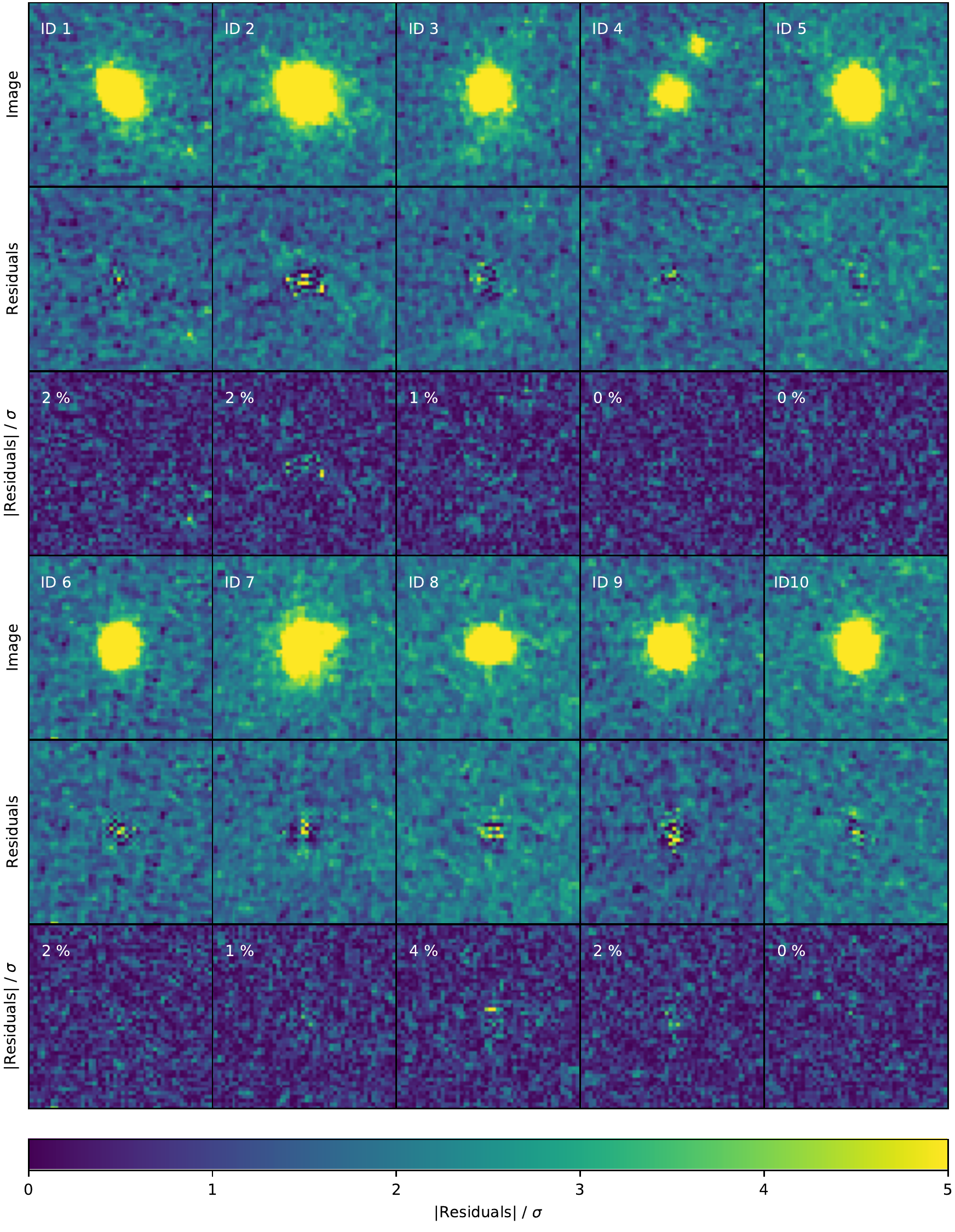}
	 \caption{\ac{hst} F160W images of the targets, residuals and their significance, defined as $|\textrm{residuals}|/\sigma$. For each source the fraction of pixels associated with the sources that have $|\textrm{residuals}|/\sigma > 3$ is indicated.}
	 \label{fig:residuals_significance}
\end{figure*}

In Figure~\ref{fig:residuals_significance} we show the F160W images and residuals after subtracting the best fit \sersic\ profiles (see Figure~\ref{fig:galfits} and Section~\ref{sec:morphology_fit}) together with a plot of the significance of the residuals that we define as the absolute value of the residuals divided by the noise in each pixel. Considering only pixels of the F160W images with a flux higher than 3 times the root mean square of the background we find that the fraction of pixels in the residual images with a significance higher than 3 is $\lesssim\SI{2}{percent}$ for all sources except for ID 8, where we find $\SI{4}{percent}$.

\bsp	
\label{lastpage}
\end{document}